\documentclass[a4paper,11pt]{article}

\usepackage{jcappub}
\usepackage{tikz}

\usepackage{color}
\usepackage{colortbl}
\definecolor{summersky}{cmyk}{0.71,0.33,0,0.14}
\definecolor{flamingo}{cmyk}{0,0.51,0.71,0.14}
\definecolor{rp}{cmyk}{0.2, 1, 0.6, 0}
\definecolor{pacificblue}{cmyk}{0.95,0.3,0, 0.19}

\begin{document}

\title{\bf $\delta M$ Formalism: A New Approach to Cosmological Perturbation Theory in Anisotropic Inflation}

\author[a]{A.~Talebian-Ashkezari,}
\author[a]{N.~Ahmadi,}
\author[b,c]{A.A.~Abolhasani}

\affiliation[a]{Department of Physics, University of Tehran, \\ Kargar Ave. North, Tehran 14395-547, Iran.}
\affiliation[b]{Department of Physics, Sharif University of Technology\\Azadi Ave., Tehran 11155-9161, Iran.}
\affiliation[c]{School of Physics, Institute for Research in Fundamental Sciences (IPM)\\Farmanieh, Tehran 19395-5531, Iran.}

\emailAdd{atalebian@ut.ac.ir}
\emailAdd{nahmadi@ut.ac.ir}
\emailAdd{abolhasani@ipm.ir}

\date{\today}

\abstract{We study the evolution of the metric perturbations in a Bianchi background in the long-wavelength limit. By applying the gradient expansion to the equations of motion we exhibit a generalized ``Separate Universe" approach to the cosmological perturbation theory. Having found this consistent separate universe picture, we introduce the $\delta M $ formalism for calculating the evolution of the linear tensor perturbations in anisotropic inflation models in {\it almost} the same way that the so-called $\delta N$ formula is applied to the super-horizon dynamics of the curvature perturbations. Similar to her twin formula, $\delta N$, this new method can substantially reduce the amount of calculations related to the evolution of tensor modes. However, it is not as general as $\delta N$; it is a "perturbative" formula and solves the shear only to linear order. In other words, it is restricted to weak shear limit.}
%
%
\maketitle
\section{Introduction and Motivation}

Cosmological perturbation theory is a pivotal step in finding the predictions of the early Universe models, e.g. inflation \cite{Inflation}. The success of inflationary paradigm can be addressed from three aspects, even if the linear order perturbations are considered. At the classical level, inflating background can tell us why has the early Universe been so flat and homogeneous. At the quantum level, inflaton vacuum fluctuations (the only thing that could be survived from inflation era) can explain the presence of very tiny initial inhomogeneities, as indispensable primordial seeds for large scale structure of the Universe. At the statistical point of view (the only way for checking theory by data in cosmology) simplest inflationary models predict nearly scale invariant, adiabatic and almost Gaussian statistics almost consistent with recent observations \cite{PLANCK}. Nevertheless, it is crucially important to go beyond the linear order to be able to discriminate among different cosmological models. For example, any tiny detection of the non-Gaussianity would rule out all the slow-roll inflationary models, since they predict non-Gaussianity of the order of the slow-roll parameters \cite{Maldacena:2002vr}.

Amid different ways for studying Einstein equations, approximation methods are important tools in cosmology and specially in the analysis of the observed anisotropies of the Cosmic Microwave Background (CMB). A powerful approximation technique used in cosmology is the so called ``long-wavelength approximation scheme'' or ``gradient expansion''. This method has been brought up and studied by many authors in cosmology previously
\cite{Lifshitz:1963ps,Belinsky:1982pk,Tomita:1972,Tomita:1975kj,Salopek:1990jq,Comer:1994np,Deruelle:1994iz,GE:2,Sasaki:1998ug,GE:III,Tanaka}. The ``quasi-isotropic'' solution of Lifshitz and Khalatnikov \cite{Lifshitz:1963ps,Belinsky:1982pk} who studied the general behavior of the space-time near the cosmological singularity, the ``anti-Newtonian'' solution of Tomita \cite{Tomita:1972,Tomita:1975kj} who investigated cosmological perturbations on super-horizon scale, and the ``long-wavelength iteration scheme'' of Salopek and Bond \cite{Salopek:1990jq} all had set up on the same approximation idea, gradient expansion. Comer et al. studied the solution of Einstein equation expanded by spatial gradient via adopting the synchronous time slice \cite{Comer:1994np}. Nonlinear perturbations near the cosmological singularity was also investigated by Deruelle and Langlois \cite{Deruelle:1994iz}.

The gradient expansion scheme employs an expansion in powers of gradient operator. It is practically built upon neglecting the inhomogeneities varying over scales smaller than a smoothing scale. When the expansion is applied to the Einstein equations of motion, as the expansion parameter goes to zero, one gets a universe locally similar to the background. One important advantage of using this method is the fact that nonlinear characteristics of the dynamical equations governing the remaining perturbations are preserved. Furthermore, at first order in parameter expansion, the perturbed dynamic equations have exactly the same form as homogeneous background. In other words, the information about perturbed fields can be found through a simple ``rescaling of the background'' fields up to higher order corrections in gradient expansion. This is the essence of the so-called ``Separate Universe'' approach \cite{Wands:2000dp,Sasaki:1998ug,Lyth:2003im}. With regards to this point, $\delta N$ formalism has been developed for computing super-horizon curvature perturbations in the context of inflationary paradigm \cite{Wands:2000dp,DeltaN:II,Naruko:2012fe,Sugiyama:2012tj}. In this formalism, the long-wavelength scalar perturbations show resemblance to (; so can be absorbed in) the integrated expansion of the background geometry evaluated from some initial time $t_0$, when all relevant fields are sufficiently outside their horizon up to the end of inflation. 

In this work, we proceed to more general background geometries for the purpose of promoting the separate universe picture and to accommodate gravitational perturbations. Therefore, for this purpose, we follow the steps below:
\begin{itemize}
\item[$\bullet$]
First of all we study the long-wavelength perturbations in Bianchi background in the 3+1, Arnowitt-Deser-Misner (ADM) decomposition \cite{Arnowitt:1962hi,Gourgoulhon_book}. This decomposition has been found to be more appropriate for the purpose of applying gradient expansion \cite{Sugiyama:2012tj,Lyth:2004gb,Tanaka,Naruko:2012fe,Comer:1996du}. 
\item[$\bullet$]
Secondly, we apply the gradient expansion to the equations of motion and come up with a set of equations valid up to the first order of the gradient expansion.
 This set of equations are the pillars of the separate universe picture of the anisotropic inflation models. 
\item[$\bullet$]
For the third step, we exhibit the consistent separate universe  picture of the cosmological perturbation theory in the Bianchi background. We show how the similarity of the long wavelength perturbations to a background parameters changes the associated parameters. Particularly we demonstrate that the perturbed equations can be recast exactly in the form of background equations. From theoretical point of view without concerning about possible applications, we study the evolution of the super-horizon metric perturbations in a Bianchi background. 
\item[$\bullet$]
The main step of this work is a generalization of the the so-called $\delta N$ formula, $\delta N = \psi$,  to a relation applicable for tensorial degrees of freedom.  Inspired by the idea of the $\delta N$ formalism, one finds that the answer lies within the geometrical shear, $\sigma_{i j}$ and anisotropic expansion of the perturbed Bianchi metric.  The simplest models with non-vanishing shear are Bianchi space-times; this is essentially the reason of studying cosmological perturbation theory in anisotropic models. To put it another way, the Bianchi background is the simplest extension of the Friedmann-Lema\^{\i}tre-Robertson-Walker (FLRW) background capable of incorporating or admitting the tensor perturbations. 
\item[$\bullet$]
Finally, we identify the (scalar and tensor) observable perturbations in the homogeneous background. Accordingly we exhibit the $\delta M$ formalism as a prescription for calculating non-trivial, linear tensorial modes in the similar fashion as the so called $\delta N$ formalism. 
\end{itemize}

For the sake of clarity, we emphasize that the constant long-wavelength gravitational waves are not physical degrees of freedom in the sense that it can be gauged away at leading order in gradient expansion via a large gauge transformation likewise adiabatic scalar perturbations \cite{Bordin:2016ruc,Maldacena:2002vr}. This implies that the only meaningful tensor perturbations on super-horizon scales are the ``would-be" decaying modes which are the ones which correspond to the shear in an anisotropic universe. These modes always fade away in an expanding background unless they are sourced by non-negligible anisotropic stress. In other words, a FLRW space-time does not have any appropriate background dynamical quantity supporting the tensorial degrees of freedom. However, in the anisotropic inflation models \cite{Aniso:Inflation,Pitrou:2008gk,Pereira:2007yy,Gumrukcuoglu:2007bx}, usually the anisotropic stress show up and as a result there are a bunch of non-trivial interactions between scalar and tensor modes. This actually makes the calculations of the correlation function a cumbersome project. By employing the proposed $\delta M$ method, the calculations of the correlation functions of the perturbations astonishingly shrink. 

Several authors have studied the behavior of gravitational waves in Bianchi-{\it I} universe \cite{GW:BIANCHI,Gumrukcuoglu:2007bx}. It is worth emphasizing that this work is \textit{not} aimed  to study the problem of the gravitational waves in a general Bianchi-{\it I} background. Instead, we exploit the Bianchi background as a suitable choice capable of incorporating tensorial degrees of freedom. This suggests that the long-wavelength tensor perturbations would redefine the integrated shear of the background metric.

In standard model of cosmology, the geometry of our Universe, smoothed on large enough scales, is well described by a spatially expanding FLRW solution. The extra degrees of freedom that control the expansion may trigger a phase of anisotropic expansion (through an anisotropic tensor). The current CMB observations show that the deviation from isotropy is small \cite{Saadeh:2016sak,PLANCK}. Therefore, in a gradient expansion approach, only the ``weak shear limit'' \cite{Pontzen:2010eg,Pitrou:2015iya} in which the induced shear is small can provide good insight into the super-horizon perturbations. In order to implement this approximation, we employ a two parameter perturbation scheme \cite{Bruni:2002sma,Sopuerta:2003rg} in which besides perturbations the geometrical shear is also considered as an extra perturbative degree. Therefore in deriving the $\delta M$ formula, two expansion schemes are applied; gradient expansion, for studying long-wavelength perturbations and weak shear limit, as a limit in which the modes evolve independently and reach to FLRW Universe. A decomposition of spatial fields into scalar, vector and tensor modes lets us to identify two independent degrees of freedom which geometrically match gravitational waves ( or vector perturbations) in FLRW space-times. Our result confirms the independent evolution of these modes at linear order.

In this paper we adopt a $(-+++)$ metric signature, and will use the Greek letters $(\mu,\nu,\alpha,\beta,...= 0,1,2,3)$ and the middle Latin indices $(i,j,k,l,...=1,2,3)$ to denote space-time indices (base space) and its spatial part, respectively.

The rest of the paper is organized as follows: In Sec. \ref{BackGeo} we study an anisotropic model of cosmology which provides a homogeneous set up in separate universe approach. In Sec. \ref{PerGeo} we study the non-linearly perturbed Einstein equations and apply the gradient expansion to those equations. In Sec. \ref{SU}, we show that separate universe picture holds in an anisotropic universe, and discuss about the dynamics of super-horizon perturbations in FLRW limit in Sec. \ref{FlRW_limit}. A short review of $\delta N$ formula is followed by the derivation of $\delta M$ formula in Sec. \ref{deltaNM}. The conclusion and discussions are given in Sec. \ref{discussions}. Some technical details are relegated into Appendices.

\section{Background Geometry}
\label{BackGeo}

The homogeneous backgrounds can be classified by general Bianchi metrics. Among those, the simplest choice with non-zero shear is Bianchi type-{\it I}
\begin{align}
\mathrm{d}s^2 &=\bar{g}_{\mu\nu}\mathrm{d}x^\mu\,\mathrm{d}x^\nu=-\mathrm{d}t^2+ \bar{\gamma}_{ij}(t)\mathrm{d}x^i\,\mathrm{d}x^j \, .
\label{backmetric}
\end{align}
It is well known that spatial part of the metric can be cast into the following form \cite{Misner:1967uu,Hawking:1968zw}:
\begin{align}
\bar{\gamma}_{ij}(t) &=e^{2N(t)}\hat{\bar{\gamma}}_{ij} \, ,\qquad \hat{\bar{\gamma}}_{ij}=(e^{2\boldsymbol{M}(t)})_{ij} \, ,
\end{align} 
where $N$ and $\boldsymbol{M}$ represent shape-preserving volume expansion and volume-preserving shape deformation, respectively. It is clear that $N$ is a scalar function of time and $\boldsymbol{M}$ is a symmetric trace free, $3 \times 3$ matrix. The matrix exponential is the series $\sum_{n=0}^{\infty}(2\boldsymbol{M})^n/n!$. Furthermore since $\hat{\bar{\gamma}}_{ij}$ is a tensor density with unit-determinant, $\boldsymbol{M}$ has to be traceless. Here and from now on, a bar over any quantity denotes its unperturbed value (unless otherwise is specified), and the indices of $M_{ij}$ will be raised by $\delta^{ij}$ and $M_{ij}$ is a measure of the homogeneous changes in Euclidean metric; so $M_{ij}=M^{i}_{j}=M^{ij}$. 

Neither $N$ nor $\boldsymbol{M}$ has any absolute geometrical meaning at a single instant of time. In the sense that, one can eliminate them at a specific time by a constant general linear transformation. However the time derivative of these quantities are of great physical significance which will be discussed extensively in the following. As an example, for a unperturbed flat FLRW space-time, $N$ would be integrated expansion $N=\int H \, \mathrm{d}t$ while one has $M_{ij}=0$, where $H=\dot{a}(t)/a(t)$ is Hubble expansion rate and $a(t)$ denotes the scale factor. For an inflationary universe $N$ is clearly the so-called total number of e-folds, $N = \ln{a(t)}$.

In the same manner as flat FLRW space-times, one can define an average Hubble expansion rate, $H$, for Bianchi space-times, as a characteristic parameter. However contrary to isotropic FLRW space-times, in Bianchi models, cosmic shear, $\hat{\bar{\sigma}}_{ij}$, as a measure of anisotropic expansion exists. The average Hubble rate and cosmic shear rate are defined in terms of time derivatives of $N$ and $M_{ij}$ as   
\begin{align}
H &\equiv \dot{N} \, ,\\
\hat{\bar{\sigma}}_{ij} &\equiv \frac{1}{2}\dot{\hat{\bar{\gamma}}}_{ij}=\frac{1}{2}\dfrac{\mathrm{d}}{\mathrm{d}t}(e^{2\boldsymbol{M}})_{ij} \, .
\end{align}
Hence $N$ can be interpreted as the integrated expansion,
\begin{align}
N=\int H \, \mathrm{d}t \, .
\end{align}
Under the assumption that $\boldsymbol{M}$ commutes with its derivative, $[\boldsymbol{\dot{M}},\boldsymbol{M}]=0$ ( See App. \ref{communication} for more details), the shear rate simply reduces to the time derivative of $M_{ij}$, as
\begin{align}
\hat{\bar{\sigma}}^{i}_{j}=\dot{M}^{i}_{j} \, .
\end{align}
Here the indices of $\hat{\bar{\sigma}}_{ij}$ are raised by $\hat{\bar{\gamma}}^{ij}=(e^{-2\boldsymbol{M}})^{ij}$ . Therefore $M_{ij}$ can be interpreted as integrated shear,
\begin{align}
M^{i}_{j}=\int \hat{\bar{\sigma}}^{i}_{j} \, \mathrm{d}t \, .
\end{align}

Applying $(3+1)$-decomposition on the metric (\ref{backmetric}), the components of $\boldsymbol{\bar{n}}$, the unit time-like vector normal to the constant cosmic time hyper-surface $\boldsymbol{\Sigma}_{t}$, will be
\begin{align}
\bar{n}^{\mu} = \left[ 1, \vec{0}\right] \, , \qquad \bar{n}_{\mu} =\left[ -1, \vec{0}\right] \, .
\end{align}
The expansion rate of the $\boldsymbol{\bar{n}}$ congruence is given by
\begin{eqnarray}
\bar{\Theta}&=& 3\dot{N} \, ,
\label{BG_expansion}
\end{eqnarray}
and its shear rate $\bar{\sigma}_{ij}$ by
\begin{eqnarray}
\bar{\sigma}_{ij}&=& e^{2N}\hat{\bar{\sigma}}_{ij} \nonumber\\
&=& \frac{1}{2}e^{2N}~\dfrac{\mathrm{d}}{\mathrm{d}t}(e^{2\boldsymbol{M}})_{ij} \, .
\label{BG_shear}
\end{eqnarray}
Note that indices of $\bar{\sigma}_{ij}$ can be raised by $\bar{\gamma}^{ij}=e^{-2N}\hat{\bar{\gamma}}^{ij}$, therefore $\bar{\sigma}^{i}_{j}=\hat{\bar{\sigma}}^{i}_{j}=\dot{M}^{i}_{j} \,$.

\subsection{Background Equations} 
In App. \ref{conf3+1}, we discuss about the Einstein equations in terms of the variables of the ADM formalism. In this subsection we study Eqs. (\ref{const.eq1.hat})-(\ref{dyn.eq4.hat}) and continuity Eqs. \eqref{continuity_E}-\eqref{continuity_J}  in the {\it background} space-time \eqref{backmetric}. In cosmic time, $t$, The only non-zero Christoffel symbols of (\ref{backmetric})  are
\begin{align}
\bar{\Gamma}^{0}_{ij}=e^{2N}[\dot{N}\hat{\bar{\gamma}}_{ij}+\hat{\bar{\sigma}}_{ij}] \, , \quad
\bar{\Gamma}^{i}_{0j}=e^{2N}[\dot{N}\delta^{i}_{j}+\hat{\bar{\sigma}}^{i}_{j}] \, .
\label{back_Cristoffel}
\end{align}
The extrinsic curvature of this hyper-surface, $K_{ij}=-\nabla_{i}\bar{n}_{j} $, is usually decomposed into trace part, $K$, and trace free part, $\hat{A}_{ij}$, as
\begin{align}
K_{ij}=\dfrac{1}{3} K \bar{\gamma}_{ij}+e^{2N}\hat{A}_{ij} \, .
\end{align}
Using Eqs. \eqref{K} and \eqref{A} we get
\begin{align}
K&=-3\dot{N} \, ,\label{kn}\\
\hat{A}_{ij}&=-\hat{\bar{\sigma}}_{ij}=-\frac{1}{2}\dot{\hat{\bar{\gamma}}}_{ij}=-\frac{1}{2}\dfrac{\mathrm{d}}{\mathrm{d}t}(e^{2\boldsymbol{M}})_{ij} \, .
\label{Am}
\end{align}

Plugging a homogeneous energy momentum tensor $T_{ij}(t)$ in Hamiltonian constraint, evolution equations for components of extrinsic curvature, $K$  and $\hat{A}_{ij}$, are given by
\begin{align}
\frac{2}{3}K^2&=\frac{2}{M^{2}_{P}} E+\hat{A}_{ij}\hat{A}^{ij}\,,
\label{const.eq1.BG}\\
\dot{K} &=\hat{A}_{ij}\hat{A}^{ij}+\frac{1}{3}K^2+\frac{1}{2M^{2}_{P}}(E+S)\,,
\label{dyn.eq2.BG}\\
\dot{\hat{A}}_{ij}&= -\frac{1}{M^2_{P}}\hat{S}_{ij} +K\hat{A}_{ij}-2\hat{A}_{ik}\hat{A}^{k}_{j} \, .
\label{dyn.eq4.BG}
\end{align}
Here $E=T_{00}$ is energy density, $S$ and $\hat{S}_{ij}$ are trace and trace free part of stress tensor, respectively. Moreover, the momentum density $J_{i}=T_{0i}$, vanishes from the momentum constraint \eqref{const.eq2.hat}. (for details see App. \ref{conf3+1}). Also the energy conservation equation \eqref{continuity_E} is given by
\begin{align}
\dot{E}+3\dot{N}(E+\dfrac{1}{3}S)+\hat{\bar{\sigma}}_{ij}\hat{S}^{ij}=0 \, .
\label{E_cconserv_background}
\end{align}
Finally, by using Eqs. \eqref{kn} and \eqref{Am}, Eq. \eqref{dyn.eq4.BG} can be rewritten as
\begin{equation}
\dot{\hat{\bar{\sigma}}}^{i}_{j} + 3 \dot{N} \hat{\bar{\sigma}}^{i}_{j} = \frac{1}{M^2_{P}}\hat{S}^{i}_{j}.
\label{EoM-sigma}
\end{equation}
which are governing equations for the dynamics of the shear tensor. 


\subsubsection{Fluid and Scalar Field Description }
\label{BG_Fluid_Scalar}
If matter can be described as a fluid, we consider the following form of the background energy momentum tensor
\begin{align}
\bar{T}_{\mu \nu} = (\bar{\rho} + \bar{p})\,\bar{u}_{\mu}\,\bar{u}_{\nu}+ \bar{p}\,\bar{g}_{\mu \nu}+ \bar{\pi}_{\mu \nu}\,.
\label{EMT_BG}
\end{align}
As we will note in App. \ref{conf3+1}, $\bar{\rho}$, $\bar{p}$ and $\bar{\pi}_{\mu \nu} $ are the relativistic energy density, the isotropic and the anisotropic pressure, respectively. Moreover momentum constraint dictates that heat transfer must vanish. $\bar{u}^\mu$ is the fluid's four-vector velocity, which its components in the comoving coordinate associated with the fluid read as
\begin{align}
\bar{u}^{\mu} = \left[ 1, \vec{0}\right] \,, \qquad \bar{u}_{\mu} =\left[ -1, \vec{0}\right] \,.
\label{back_u}
\end{align}
In this case fluid's velocity is not tilted relative to the normal vector of constant time hyper-surface, i.e. $\boldsymbol{\bar{n}}=\boldsymbol{\bar{u}}$. It is easily to check that $E=\rho$, $J_{i}=0$, $S=3p$ and $\hat{S}_{ij}=e^{-2N}\bar{\pi}_{ij} \equiv \hat{\bar{\pi}}_{ij}$.

By using (\ref{back_Cristoffel}), (\ref{EMT_BG}) and (\ref{back_u}), the energy conservation equation, $-u_{\mu} \nabla_{\nu} T^{\mu \nu} = 0$, can be written as
\begin{align}
\label{continuityBG}
\dot{\bar{\rho}} + 3 \dot{N} (\bar{\rho} + \bar{p}) + \hat{\bar{\sigma}}_{ij}\hat{\bar{\pi}}^{ij}=0 \,.
\end{align}

It is worth mentioning that the evolution of a scalar field (\ref{EoMfi}) in this background is given by the following equation
\begin{equation}
\left(\sqrt{\bar{\gamma}} \dot{\bar{\varphi}} \right)\dot{} + \sqrt{\bar{\gamma}}~V_{\varphi} = 0 \,,
\label{EoMBGfi}
\end{equation}
where $\bar{\gamma}=\det[\bar{\gamma}_{ij}]=e^{6N}$.

\section{Perturbed Geometry}
\label{PerGeo}

Having calculated background equation of a Bianchi-{\it I} universe, it is time to tackle the problem in the presence of the perturbations. We write the perturbed metric in the ADM ( 3+1 decomposition) form
\begin{align}
\mathrm{d}s^2 &= g_{\mu\nu}\mathrm{d}x^{\mu}\mathrm{d}x^{\nu} \nonumber\\
&=-\alpha^2(t,\mathbf{x})~\mathrm{d}t^2+\gamma_{ij}(t,\mathbf{x})\Big(\mathrm{d}x^i+\beta^{i}(t,\mathbf{x})~\mathrm{d}t\Big)\Big(\mathrm{d}x^j+\beta^{j}(t,\mathbf{x})~\mathrm{d}t\Big) \, ,
\label{ADMmetric}
\end{align}
where $\alpha$ and $\beta^i$ are lapse function and shift vector, respectively in ADM approach and $\boldsymbol{\gamma}$ is the spatial metric of constant-$t$ hyper-surfaces, $g_{ij}=\gamma_{ij}$.
The spatial indices can be raised and lowered by $\gamma_{ij}$, e.g. $\beta_{i}=\gamma_{ij}\beta^{j}$. We consider the following decomposition for the perturbed spatial metric 
\begin{align}
\gamma_{ij}(t,\mathbf{x}) &=e^{2\mathcal{N}(t,\mathbf{x})}(e^{2\boldsymbol{\mathcal{M}}(t,\mathbf{x})})_{ij} \, ,
\label{pertgamma2}
\end{align}
where $\mathcal{N}$ is the scalar function and $\mathcal{M}_{ij}$ is a traceless $3 \times 3$ matrix. Then, we separate the homogeneous and inhomogeneous parts of $\mathcal{N}$ and $\mathcal{M}_{ij}$ in this way 
\begin{align}
\mathcal{N}(t,\mathbf{x}) &=N(t)+\psi(t,\mathbf{x}) \, ,\label{N_psi}\\
\mathcal{M}_{ij}(t,\mathbf{x}) &=M_{ij}(t)+h_{ij}(t,\mathbf{x}) \, ,\label{M-h}
\end{align}
in which $\psi$ and $h_{ij}$ are metric perturbations. Similar to FLRW background, $\psi$ quantifies the curvature perturbations. $h_{ij}$ may be represented by $3 \times 3$ symmetric traceless matrix which can be decomposed into scalar, vector and tensor modes in Bianchi-I backgrounds \cite{Pereira:2007yy,Pitrou:2008gk}.

In the rest of this section, we discuss about the evolution equations in long-wavelength perturbations limit. As it is mentioned in introduction, a successful approach to study these perturbations, would be the gradient expansion method.
\subsection{First Expansion Scheme: Gradient Expansion}
\label{gradient_expansion}
The gradient expansion for any quantity is an expansion in the spatial gradient, $\partial_{k}$, of the quantity. In practice, after focusing on a fixed time, each spatial gradient is multiplied by a parameter $\epsilon$, and then the expansion as a power series in $\epsilon$ is performed. The expansion would be useful if relevant quantities are assumed to be smooth on a smoothing scale, which we denote it by $L$. In cosmological applications, this scale is chosen to be somewhat below the shortest scale of cosmological interest, $L<k^{-1}$. To find the smoothing scale for the perturbed universe described above, one notes that

at each point $\mathbf{x}$ on a spatial hyper-surface with the spatial three metric $\gamma_{ij}(t,\mathbf{x})$, there are couple of characteristic scales. We denote a characteristic comoving length in $x^k$ direction, i.e., the scale of inhomogeneity by $L_k$,
\begin{align}
\partial_{k}\gamma_{ij} \thicksim L_{k}^{-1}\gamma_{ij} \,,
\label{space_L}
\end{align}
and introduce $L$, as the {\it biggest comoving scale} among the set of $L_k$s and other scales introduced  by the shear eigenvectors or energy momentum tensor. By virtue of the first assumption in a separate universe model, all the scales which are small compared to $L$ are smoothed. In other words, on lengthscales smaller than $L$, the space is almost homogeneous and the 3-metric depends only on time. It is worth emphasizing that on these scales the metric is homogeneous but not necessarily isotropic \cite{Dimopoulos:2008yv}. A second distinguished scale, $T^{-1}$ is the {\it biggest comoving scale}\footnote{This fast timescale corresponds to the timescale of interaction of anisotropy source with metric and is equal to that of non-adiabatic (vector/ tensor) metric perturbations, while for adiabatic ones, the timescale is $\left(aH\right)^{-1}$.} on which the background 3-metric evolves,
\begin{align}
\partial_{t}\gamma_{ij} \thicksim T \gamma_{ij} \,.
\label{time_H}
\end{align}
In a FRW background, $\left(aH\right)^{-1}$ is the only geometric scale, where $H=\dfrac{\dot{a}}{a}$ and $a=(\det[\gamma_{ij}])^{1/6}$. In Bianchi spacetimes, however, three scales are introduced by the shear eigenvectors and one more by energy momentum tensor. Assuming that those are all of the same order, we define a parameter $m$ which relates different scales together, $\left(aH\right)^{-1}=mT^{-1}$. We implicitly presumed that $m<1$.
In the scales shorter than the smoothing length scale $L$, one can safely ignore all the spatial gradient terms with respect to the time derivative terms:
\begin{align}
\partial_{k}\gamma_{ij} \ll \partial_{t}\gamma_{ij}
~~ \Longleftrightarrow ~~ L \gg T^{-1} \,.
\label{GE_condition}
\end{align}
It is easy to see that when (\ref{GE_condition}) is satisfied, $L\gg \left(aH\right)^{-1}$ would be trivially valid. We also suppose that the sphere of comoving radius $L$ does not affect on a cosmological scale $k^{-1}$, although it greatly simplifies the analysis, as we shall see. In the language of perturbation theory, $L^{-1}$ is related to the comoving wavenumber of perturbations, $k$. So we can define gradient expansion parameter as 
\begin{align}
\epsilon \equiv \dfrac{k}{T} \,.
\end{align}

With regard to (\ref{GE_condition}), we have $\epsilon \ll 1$, so in any gradient expansion, one can omit higher order terms in $\epsilon$ to the desired accuracy. When applied to Einstein equations of motion, at the leading order in $\epsilon$ the universe locally evolves like the homogeneous background universe with appropriate modified parameters. In the very long-wavelength limit, say $L \rightarrow \infty$, all the perturbations are ignored and our universe is assumed to be smooth. By this assumption, the universe at each position will evolve as if it were homogeneous, which means the fields will obey evolution equations like those in a homogeneous universe, but now described by {\it local} scale factor, Hubble expansion rate and shear tensor. This approach is the separate universe approach to the cosmological perturbation theory. In the separate universe picture there exists an appropriate set of coordinates in which the metric and equations of motions of any local region can be written as the unperturbed background form.
\subsection{Gradient Expansion of Perturbed Equations}
\label{GE}
Since we are interested in the evolution of long-wavelength perturbations, we shall apply gradient expansion on the field equations. Our physical assumption is that the smoothing length is sufficiently large, i.e. $\epsilon \rightarrow 0$. Then all the perturbative quantities 
 must disappear in this limit and the universe becomes locally a Bianchi one. To be precise, by locally we mean, a region sufficiently larger than the new horizon, $T^{-1}$, but significantly smaller than the smoothing scale $L$. Let $s$ be a comoving scale in the region, we have $T^{-1}<s<L< k^{-1}$. Here, and throughout, superhorizon perturbations refer to those that satisfy $k \ll T$; so the Bianchi universe approximation is valid throughout the entire superhorizon era, $k \ll T$.

There are two classes of perturbations. First, those who have background {\textit{counterparts}}.
The second class, there are perturbations which are {\textit{not supported}} by the background dynamics. these quantities must vanish by going to the smoothing limit, $\epsilon \rightarrow 0$. In this subsection we shall show that $\psi$, $h_{ij}$, $\delta E$, $\delta S$, and $\delta \hat{S}_{ij}$ correspond to the first class, whilst $\beta_{i}$ and $\mathcal{J}_{i}$ must vanish in this limit, namely as $\beta_{i}\sim \mathcal{O}(\epsilon^{n})$, and $\mathcal{J}_{i}\sim \mathcal{O}(\epsilon^{n'})$. Here, $\mathcal{J}_{i} = T_{\mu \nu} n^{\mu} \Sigma^{\nu}_{i}$ refers to the perturbed momentum density. Also note that $\alpha$ is not constrained by separate universe assumptions.
It is actually of order $\sim \epsilon^{0}$. In studying the perturbed equations, we will work with $\tau$ instead of $t$, as local proper time. 

Let us examine what the momentum constraint tells us about the gradient expansion ordering of $\beta_{i}$ and $\mathcal{J}_{i}$. Plugging
\begin{align}
\mathcal{J}_{i}=\dfrac{1}{\alpha}(T_{i0}-\beta^{k} T_{ki})
\end{align}
into Eq. \eqref{const.eq2.hat} yields
\begin{align}
D_{j}\hat{\mathcal{A}}^{j}_{i}-\frac{2}{3}\partial_{i}\mathcal{K}&=\frac{1}{M^{2}_{P}}\dfrac{1}{\alpha}(\beta^{k} T_{ki}-T_{i0}) \,.
\label{const.eq2.PG_J}
\end{align}
Hence, the momentum constraint can be satisfied if the magnitudes of shift vector and $T_{i0}$ are at most of the first order in the parameter of gradient expansion, say
\begin{align}
\mathcal{J}_{i} \sim T_{i0} \sim \beta_{i} \sim \mathcal{O}(\epsilon)\,.
\label{J_beta_Ti0}
\end{align}

This result would be our guideline to apply gradient expansion to the field equations. Now, by using \eqref{K} and \eqref{A} and noting that $\beta_{i} \sim \mathcal{O}(\epsilon)$, trace and traceless parts of the extrinsic curvature, will be
\begin{align}
\mathcal{K}=-3{\mathcal{N}}'+\mathcal{O}(\epsilon^2) \,, \quad \hat{\mathcal{A}}_{ij}=-\hat{\sigma}_{ij}+\mathcal{O}(\epsilon^2) \,,
\label{KN_AM}
\end{align}
where
\begin{align}
\hat{\sigma}_{ij}=\frac{1}{2}\hat{\gamma}'_{ij} \,, \quad \hat{\gamma}_{ij}=(e^{2\boldsymbol{\mathcal{M}}})_{ij} \,.
\label{sigma_M}
\end{align}
Here, and throughout, the primes indicate derivative with respect to proper time, $\tau$,
\begin{align}
\mathrm{d}\tau=(\alpha^2-\beta^{k}\beta_{k})^{1/2}~\mathrm{d}t=\alpha~\mathrm{d}t+\mathcal{O}(\epsilon^2) \,.
\end{align}
Using a general energy momentum tensor $T_{\mu \nu}(t,\mathbf{x})$, ADM equations (\ref{const.eq1.hat}),(\ref{dyn.eq1.hat})-(\ref{dyn.eq4.hat}) read as:
\begin{align}
\frac{2}{3}\mathcal{K}^2&=\frac{2}{M^{2}_{P}} \mathcal{E}+\hat{\mathcal{A}}_{ij}\hat{\mathcal{A}}^{ij}+\mathcal{O}(\epsilon^2) \, ,
\label{const.eq1.PG}\\
\mathcal{K}' &=\hat{\mathcal{A}}_{ij}\hat{\mathcal{A}}^{ij}+\frac{1}{3}\mathcal{K}^2+\frac{1}{2M^{2}_{P}}(\mathcal{E}+\mathcal{S})+\mathcal{O}(\epsilon^2) \, ,
\label{dyn.eq2.PG}\\
\hat{\mathcal{A}}'_{ij}&= -\frac{1}{M^2_{P}}\hat{\mathcal{S}}_{ij} +\mathcal{K}\hat{\mathcal{A}}_{ij}-2\hat{\mathcal{A}}_{ik}\hat{\mathcal{A}}^{k}_{j}+\mathcal{O}(\epsilon^2) \, .
\label{dyn.eq4.PG}
\end{align}

The above relations form a complete set of equations for studying the dynamics of a smoothed patch of the Universe. Also the continuity equation \eqref{continuity_E} is given by
\begin{align}
\mathcal{E}' + 3 \mathcal{N}' (\mathcal{E}+\dfrac{1}{3}\mathcal{S}) + \hat{\sigma}_{ij}\hat{\mathcal{S}}^{ij}=\mathcal{O}(\epsilon^2) \,.
\label{E_conserv}
\end{align}

\subsubsection{Fluid and Scalar Field Description }
\label{PG_Fluid_Scalar}
Using the definition of the energy momentum tensor for a generalized fluid \eqref{EMT}, the momentum density is given by
\begin{align}
-\mathcal{J}_{i}=q_{i}+\dfrac{1}{\alpha}(\beta^{k}+v^{k})\left( \pi_{ik}+(\rho+p)\gamma_{ik}\right)+\mathcal{O}(\beta^3,v^3,\beta^2v,\beta v^2,v^3,\beta^2q,v^2q,\beta v q) \, .
\end{align}
Therefore, one can satisfy the momentum constraint by considering that the first non-zero terms in gradient expansion of the shift vector, spatial velocity and heat transfer are
\begin{align}
\beta_{i} \sim v_{i} \sim q_{i} \sim \mathcal{O}(\epsilon)\,.
\label{beta_v_q}
\end{align}
The first relation in \eqref{matter_condition} implies that
\begin{align}
q_{0}&=-v^{i}q_{i} \sim \mathcal{O}(\epsilon^2) \, ,\nonumber\\
\pi_{00}&=v^{i}v^{j}\pi_{ij} \sim \mathcal{O}(\epsilon^2) \, ,\nonumber\\
\pi_{0i}&=-v^{j}\pi_{ij} \sim \mathcal{O}(\epsilon) \, .
\label{q0_pi} 
\end{align}

Using above gradient expansion ordering, one can see that $\mathcal{E}=\rho/\alpha^2+\mathcal{O}(\epsilon^2)$, $\mathcal{S}=3p+\mathcal{O}(\epsilon^2)$, $\hat{\mathcal{S}}_{ij}=e^{-2\mathcal{N}}\pi_{ij}+\mathcal{O}(\epsilon^2) \equiv \hat{\pi}_{ij}+\mathcal{O}(\epsilon^2)$.

It would be advantageous to apply the gradient expansion to the continuity equation and the equation of motion of a scalar field. Using \eqref{beta_v_q}, \eqref{q0_pi}, and the ordering of Christoffel symbols given in App. \ref{GEChris}, one can see that the energy is conserved up to $\mathcal{O}(\epsilon^2)$,
\begin{align}
\label{continuity}
\rho' + 3 \mathcal{N}' (\rho + p) + \hat\sigma_{ij}\hat\pi^{ij}=\mathcal{O}(\epsilon^2) \, .
\end{align}
Also, by using Cramer’s rule, $g=-\alpha^2 \gamma$, where $g=\det[g_{\mu\nu}]$ and $\gamma=\det[\gamma_{ij}]$, the equation of motion of a scalar field (\ref{EoMfi}) can be rewritten as
\begin{equation}
\left(\sqrt{\gamma} \varphi' \right)' + \sqrt{\gamma}~V_{\varphi} = \mathcal{O}(\epsilon^2) \,,
\label{EoMPGfi}
\end{equation}
where $\gamma=\det[\gamma_{ij}]=e^{6\mathcal{N}}+\mathcal{O}(\epsilon^2)$.

\section{Separate Universe}
\label{SU}

Separate universe approach to the cosmological perturbation theory claims that  perturbed universe, on scales much smaller than the wavelength of the smoothing scale, behaves locally as a separate universe with different background parameters. We will show that these regions evolve as  locally homogeneous universes independently. As it is mentioned before, in the so-called separate universe approach perturbation equations (in the long-wavelength limit) are brought back to the form of the background equations. We show that the separate universe picture is valid in the \textit{finite momentum} limit to the first order in the gradient expansion $\epsilon$. This means that the non-trivial physical perturbations can be incorporated into the background in the super-horizon limit. 

By the way, comparing the perturbed equations \eqref{const.eq1.PG}- \eqref{dyn.eq4.PG} with those of the homogeneous background \eqref{const.eq1.BG}- \eqref{dyn.eq4.BG}, the perturbation equations can be drawn out of the background equations through the following replacements,
\begin{align}
t &\rightarrow \tau (t,\mathbf{x}) \, ,\label{Tau}\\
K(t) &\rightarrow \mathcal{K}(t,\mathbf{x}) \, ,\label{K_replacement}\\
\hat{A}_{ij}(t) &\rightarrow \hat{\mathcal{A}}_{ij}(t,\mathbf{x}) \, ,\label{A_replacement}\\
E(t) &\rightarrow \mathcal{E}(t,\mathbf{x}) \, ,\\
S(t) &\rightarrow \mathcal{S}(t,\mathbf{x}) \, ,\\
\hat{S}_{ij}(t) &\rightarrow \hat{\mathcal{S}}_{ij}(t,\mathbf{x}) \, .
\end{align}
Moreover, the perturbed continuity equation \eqref{continuity} and perturbed scalar field equation \eqref{EoMPGfi} match the background versions  \eqref{continuityBG} and \eqref{EoMBGfi} by applying the above replacements. 
Let us emphasize that Eqs. \eqref{const.eq1.PG}-\eqref{dyn.eq4.PG} are valid to {\it all order} in perturbations $\delta$ with corrections up to second order of gradient expansion ${\cal O}(\epsilon^2)$. It should be noted that the gradient expansion parameters can be chosen arbitrarily small irrespective of the order of perturbations. The above results, indicate that
\begin{quote}
{\it Every super-horizon smoothed patch of the Universe (with the size much smaller than the wavelength of the perturbations) evolves like a separate Bianchi universe with its own energy density and stress tensor which varies from patch to patch.}
\end{quote}
This shows that there is a consistent separate universe picture for the case of the generalized fluid in the Bianchi geometry.

The fundamental result of the above analysis is that the ``functional form" of the evolution of each (locally) perturbed separate universe patch is the same as the background. As a result, the solution for the perturbed equation can be momentarily found by simple replacements in the background solutions. It is worth mentioning that all the information about inhomogeneities are enfolded in the initial conditions and in the proper time $\tau$. For an example, assuming the following form for the solution of the scalar field equation \eqref{EoMBGfi},
\begin{align}
\bar{\varphi}(t)&=\bar{\varphi}[t,\bar{\varphi}_{0}(t_{0}),\dot{\bar{\varphi}}_{0}(t_{0})] \, ,
\end{align}
the solution of the perturbation equation \eqref{EoMPGfi} can be found almost for free as 
\begin{align}
\varphi(\tau,\mathbf{x})&=\bar{\varphi}[\tau,\varphi_{0}(\tau_{0},\mathbf{x}),\varphi'_{0}(\tau_{0},\mathbf{x})] \, .
\end{align}
Here, $\bar{\varphi}_{0}(t_{0})$ and $\varphi_{0}(\tau_{0},\mathbf{x})$ denote initial value of scalar field solution in the background and perturbed level, respectively, and $\varphi_{0}(\tau_{0},\mathbf{x})$ is computed at position $\mathbf{x}$.

 The separate universe picture in Bianchi geometry holds for any choice of gauge which respects $\beta^i ={\cal O} (\epsilon)$ condition. One may dispute that the resemblence of the equations for homogeneous fields to the equations at gradient expansion where the spatial gradients are neglected seems trivial. The non-trivial point in our work is related to smoothing large but {\it finite} wavelengths and then studying Fourier modes of interest in superhorizon era, $k\ll T$.
 In other words, to the leading order of the gradient expansion, the separate universe picture is valid for a {\it finite} smoothing scale. 

Apart from the anisotropy of the smoothed universe, there is another subtle difference between separate universe picture in anisotropic and isotropic models. In the latter, some degrees of freedom are gauged away  
 by an appropriate coordinate transformation. In this regard, Sugiyama et.al choose the $\psi=C^i=0$ to fix the gauge completely \cite{Sugiyama:2012tj}, where $C^i$ refers to the scalar and vector part of $h_{ij}$. There are other alternative choices for fixing the gauge (see e.g. \cite{Sasaki:1998ug} and \cite{Tanaka}), too. However, the intriguing point about our work is that the smoothed universe model serve as an unperturbed (Bianch I) universe without imposing any further gauge condition.

\section{FLRW Limit}
\label{FlRW_limit}
Thus far, we have accomplished setting up a consistent separate universe picture of the Bianchi geometries. However, a realistic and observationally consistent model for the Universe is a simple FLRW model with high spatial symmetry. Our achievements would be untenable unless we study the perturbations on a near isotropic Bianchi model. This study is aimed to find tensor perturbations in near FLRW anisotropic inflationary models; so we have to decompose the metric perturbations to identify scalars, vectors and tensors  in isotropic limit. Fortunately, going to the zero-shear limit makes the calculations remarkably simple. As it is shown in the following, in this limit the evolution of different components of the shear tensor as well as the isotropic expansion decouple which is a great relief.

In this section, we first decompose the traceless, symmetric spatial 3-tensors into different components and then find the dynamics of each component in zero-shear limit. By using two parameter perturbation scheme introduced in \cite{Bruni:2002sma} and shortly reviewed in Appendix E, we will derive nearly-isotropic limit of those classes. In this section, we work with background values of $\hat{\bar{\sigma}}_{ij}$ and $\hat{\bar{\gamma}}_{ij}$, but we omit bars above them, for brevity.
 
 \subsection{Adaptive Decomposition}

For studying inhomogeneities in an anisotropic background, it is very useful to decompose the spatial perturbations of metric and matter in a local basis adapted to the given wave vector $k_{i}$ \cite{Pereira:2007yy}. One should be careful about non-trivial relation between vectors and co-vectors in anisotropic geometries. The comoving wave co-vectors are constant, $\dot{k}_{i}=0$; while the dual vectors, $k^i=\hat{\gamma}^{ij}k_{j}$, change with time.

We define a set of local, orthonormal basis $\{ \hat{k}_{i},e^{2}_{i},e^{3}_{i} \}$ which span the constant time hyper-surface. $\hat{k}^{i}$ corresponds to the normalized momentum of a mode and bases $\{ e^{2}_{i},e^{3}_{i} \}$ span the subspace orthogonal to $\hat{k}^{i}$. Note that the above basis is defined up a a rotation about the $k^i$ \footnote{
The decomposition of any (3-dimensional) vector and symmetric tensor in this basis are given by
\begin{align}
V_{i} &= (\hat{k}^{j} V_{j}) \hat{k}_{i} + P^{j}_{i} V_{j},\\
V_{ij} &= (\dfrac{1}{3} \hat{\gamma}^{kl} V_{kl}) \hat{\gamma}_{ij} + (\dfrac{3}{2} T^{kl} V_{kl}) T_{ij} + 2 \hat{k}_{(i} (P^{m}_{j)} \hat{k}^{n} V_{mn}) + \Lambda^{mn}_{ij} V_{mn}.
\end{align}
Here $P_{ij} \equiv \hat{\gamma}_{ij}-\hat{k}_{i}\hat{k}_{j}$, $T_{ij} \equiv \hat{k}_{i}\hat{k}_{j} - \dfrac{1}{3}\hat{\gamma}_{ij}$ and $\Lambda^{ab}_{ij} \equiv P^{a}_{i}P^{b}_{j}-\dfrac{1}{2}P_{ij}P^{ab} $ are the projection operator onto the subspace perpendicular to $\hat{k}^{i}$, trace extracting operator and the projector on tensor modes, respectively.}.

Any traceless symmetric spatial 3-tensor $X_{ij}$ has 5 independent components which can be decomposed on the local basis introduced above in the following way
\begin{align}
X_{ij} &= \dfrac{3}{2}(\hat{k}_{i} \hat{k}_{j}-\dfrac{1}{3}\hat{\gamma}_{ij} )X_{\Vert}
+ 2\sum_{a=2,3} \hat{k}_{(i} e^{a}_{j)}X_{a} + \sum_{\lambda=+, \times} e^{\lambda}_{ij}X_{\lambda} \, .
\label{X_decomposition}
\end{align}
The components, $\{X_{\Vert},X_{a}, X_{\lambda} \}$, can be obtained by applying projection along $\hat{k}^{i} \hat{k}^{j}$, $\hat{k}^{(i} e_{a}^{j)}$ and the 
polarization tensor $e^{\lambda}_{ij}$ as
\begin{align}
e^{\lambda}_{ij}=\dfrac{e^{2}_{i}e^{2}_{j}-e^{3}_{i}e^{3}_{j}}{\sqrt{2}}\delta^{\lambda}_{+}+\dfrac{e^{2}_{i}e^{3}_{j}+e^{3}_{i}e^{2}_{j}}{\sqrt{2}}\delta^{\lambda}_{\times} \, .
\end{align}
It can be easily checked that
\begin{align}
\hat{\gamma}_{ij}X^{ij} = 0 \, , ~~~X_{\Vert} = X_{ij} k^{i} k^{j} \, , ~~~X_{a} = X_{ij} k^{i} e_{a}^{j} \, , ~~~X_{\lambda} = X_{ij} e^{ij}_{\lambda} \, .
\label{X_components}
\end{align}

Let us emphasize that though $ X_{\Vert} ,X_{a}$ and $ X_{\lambda}$ explicitly depend on $k_i$ they should not be interpreted as the Fourier components of $X_{ij}$, since this dependence appear solely as a result of local anisotropy of space.

Now, by employing Eq. \eqref{X_components} the equations of motion \eqref{const.eq1.BG}-\eqref{E_cconserv_background} transform to more convenient forms. Exploiting Eq. \eqref{X_components} the trace of square of shear $\sigma^2 \equiv \hat{\sigma}^{ij} \hat{\sigma}_{ij}$ can be written as 
\begin{align}
\label{sigma2}
\sigma^2 \equiv \hat{\sigma}^{ij} \hat{\sigma}_{ij} = \dfrac{3}{2}\sigma_{\Vert}^2 + 2\sum_{a=2,3} \sigma_{a}^{2} + \sum_{\lambda = +,\times}\sigma_{\lambda}^{2} \, .
\end{align}
In a like manner, the mixed term that appears in continuity equation \eqref{E_cconserv_background} expanded to
\begin{align}
\label{sigma-S}
\hat{\sigma}^{ij} \hat{S}_{ij} = \dfrac{3}{2}\sigma_{\Vert} \hat{S}_{\Vert} + 2\sum_{a=2,3} \sigma_{a} \hat{S}_{a} + \sum_{\lambda = +,\times}\sigma_{\lambda} \hat{S}_{\lambda} \, .
\end{align}
 As it will be discussed in the following section the anisotropy of the observed universe put observational bounds on the different components of the shear tensor, $\sigma_{\Vert}, \sigma_{a}, \sigma_{\lambda} \ll H$. 

Now, using Eq. \eqref{sigma2} and \eqref{sigma-S}, the background equations of \eqref{const.eq1.BG},\eqref{dyn.eq2.BG} and \eqref{E_cconserv_background} can be rewritten in terms of the components of $\hat{\bar{\sigma}}_{ij}$ and $\hat{S}_{ij}$ as
\begin{align}
\dot{N}^2 - \frac{1}{3M^{2}_{P}} E &= \dfrac{1}{4}\sigma_{\Vert}^2 + \dfrac{1}{3}\sum_{a=2,3} \sigma_{a}^{2} + \dfrac{1}{6} \sum_{\lambda = +,\times}\sigma_{\lambda}^{2}\,,
\label{11}\\
\ddot{N}+\dot{N}^2 +\frac{1}{6M^{2}_{P}}(E+S) &= -\dfrac{1}{2}\sigma_{\Vert}^2 - \dfrac{2}{3}\sum_{a=2,3} \sigma_{a}^{2} -\dfrac{1}{3} \sum_{\lambda = +,\times}\sigma_{\lambda}^{2}\,,
\label{22}\\
\dot{E}+3\dot{N}(E+\dfrac{1}{3}S) &= -\dfrac{3}{2}\sigma_{\Vert} \hat{S}_{\Vert} - 2\sum_{a=2,3} \sigma_{a} \hat{S}_{a} - \sum_{\lambda = +,\times}\sigma_{\lambda} \hat{S}_{\lambda}  \,
\label{44}
\end{align}
in which the Eq.\eqref{kn} and \eqref{Am} are used. Moreover, the time evolution of components of the shear tensor $(\sigma_{\Vert}, \sigma_{a}, \sigma_{\lambda})$ can be easily obtained from Eq. \eqref{EoM-sigma} 
\begin{align}
\dot{\sigma}_{\Vert}+3\dot{N}\sigma_{\Vert}&=\dfrac{1}{M^{2}_{P}}\hat{S}_{\Vert}-2\sum_{a}\sigma^{2}_{a} \, ,\label{sigma_Vert}\\
\dot{\sigma}_{a}+3\dot{N}\sigma_{a}&=\dfrac{1}{M^{2}_{P}}\hat{S}_{a}+\dfrac{3}{2}\sigma_{a}\sigma_{\Vert}-2\sum_{b,\lambda}\sigma_{b}\sigma_{\lambda}C^{\lambda}_{ab} \, ,\label{sigma_a}\\
\dot{\sigma}_{\lambda}+3\dot{N}\sigma_{\lambda}&=\dfrac{1}{M^{2}_{P}}\hat{S}_{\lambda}+2\sum_{a,b}\sigma_{a}\sigma_{b}C^{\lambda}_{ab} \, ,\label{sigma_lambda}
\end{align}
where $C_{ab}^{\lambda}$ is defined as
\begin{align}
C_{ab}^{\lambda}=e_{a}^{i}\,e_{b}^{j}\,e^{\lambda}_{ij} \, .
\end{align}
\subsection{Second Expansion Scheme: Weak Shear Limit}
\label{Small_shear_expansion}
We recall that Bianchi universe models are anisotropic expanding models that are spatially homogeneous. These models fall into different classes; certain types of Bianchi models (Bianchi type-I for an example) have continuous isotropic limit \cite{Ellis:1968vb,Pontzen:2010eg}.

In this limit, in which the rotational symmetry is almost restored, the shear tensor can be decomposed into five ``non-interacting" modes (i.e. scalar, vector and tensor) under $SO(3)$ group \cite{Saadeh:2016sak,Pontzen:2010eg}. There are different upper limits on different modes of shear tensor. However for our purpose it is enough to work with
\begin{align}
m \sim \dfrac{\sigma_{ij}}{H},\label{m}
\end{align}
which quantifies the \textit{typical} ratio of the shear tensor modes to isotropic expansion rate $H=\dot{N}$. There is a theoretical upper bound on $m$ (and in turn on $\hat{\sigma}_{ij} \hat{\sigma}^{ij}$). The local positivity of energy density in the Hamiltonian constraint \eqref{const.eq1.BG} implies that $m < \sqrt{6}$ \cite{Pereira:2007yy}. Roughly speaking, according to the observations this ratio is limited to $m \lesssim 10^{-6}$ \cite{PLANCK,Saadeh:2016sak} .
 
With regard to the above considerations, we have\footnote{In the following, the order shown in the parantheses refer to the tensors nondimentioned with adequate powers of isotropic expansion, $H$.}
\begin{align}
M^{i}_{j} \sim \mathcal{O}(m) \, ,~~~\hat{\sigma}^{i}_{j} \sim \mathcal{O}(m) \, ,~~~\hat{S}^{i}_{j} \sim \mathcal{O}(m) \, .
\end{align}
 In order to attain the isotropic limit, we employ a two-parameter perturbation scheme. Consider a background FLRW space-time and a family of space-times  diffeomorphic to the background for describing the perturbed geometrics. Each member of the family is labelled by some value of parameters $(m,\delta)$, representing the shear and perturbation strength, respectively. Besides the usual metric and matter perturbations, the geometrical shear (and/or anisotropic stress) is also considered as an extra orthogonal perturbative degree of freedom. For any quantity $X$, one considers a perturbation series $X = \sum_{a,b = 0}^{\infty} X^{(a,b)}$ in which $(a,b)$ indicate powers in $\sigma$ and the order of (matter/metric) perturbation, respectively. In small shear approximation, $a$ is limited to be at most one. One can simply check that homogeneous perturbation, $M_{ij}$, is of order (0,1) while curvature perturbation, $\psi$, and different components of $h_{ij}$ are of order (1,1).
 More details and some rigorous mathematical considerationson about two parameter perturbation theory can be found in App.\ref{Two-parameter}.

 It is easy to see that, in the isotropic limit $m \rightarrow 0$, Eqs. \eqref{11}-\eqref{sigma_lambda} shrink to 
\begin{align}
\dot{N}^2&=\frac{1}{3M^{2}_{P}} E+\mathcal{O}(m^2)\,,
\label{1}\\
\ddot{N}+\dot{N}^2 &=-\frac{1}{6M^{2}_{P}}(E+S)+\mathcal{O}(m^2)\,,
\label{2}\\
\dot{E}+3\dot{N}(E+\dfrac{1}{3}S)&= \mathcal{O}(m^2) \, ,\label{3}\\
\dot{\sigma}_{\chi}+3\dot{N}\sigma_{\chi}&= \frac{1}{M^2_{P}}\hat{S}_{\chi} +\mathcal{O}(m^2) \, ,~~~~~~~~ \chi = \Vert,a,\lambda
\label{4}
\end{align}
These equations are the nearly-isotropic Bianchi equations, which among them Eqs.\eqref{1}-\eqref{3} are the FLRW background equations up to $\mathcal{O}(m^2)$, and \eqref{4} governs the evolution of shear components which can be used to derive the time evolution of $M_{ij}$ components. In App.\ref{communication} we show that $\hat{\sigma}^{i}_{j} = \dot{M}^{i}_{j} + \mathcal{O}(m^3)$ which in turn give the components of shear, $\sigma_{ij}$, in terms of the time derivative of the integrated shear. In the zero shear limit, one readily finds 
\begin{align}
\sigma_{\chi} &= \dot{M}_{\chi}+\mathcal{O}(m^2) \, ,~~~~~~~~~~ \chi = \Vert,a,\lambda.
\end{align}
Note that, the ${\cal O}(m^2)$ corrections would disappear in case one has $[\boldsymbol{M},\dot{\boldsymbol{M}}]=0$. The basis vectors $\{\hat{k}_{i},e^{a}_{i}\}$ and consequently the polarization tensor $e^{\lambda}_{ij}$ have non-vanishing time derivative in order to keep their orthonormal structure on each constant time hyper-surface. The time evolution of these quantities are of order $\mathcal{O}(m)$ \cite{Pereira:2007yy}. As a result, the components of anisotropies satisfy the following equations in the FLRW limit of the background geometry:
\begin{align}
\ddot{M}_{\chi}+3\dot{N}\dot{M}_{\chi}&=\dfrac{1}{M^{2}_{P}}\hat{S}_{\chi}+\mathcal{O}(m^2) \, ,~~~~~~~~~~ \chi = \Vert,a,\lambda.
\label{M_chi}
\end{align}
Having found the dynamics of homogeneous part of components of integrated shear $M_{ij}$ in the isotropic limit, the dynamics of those components in the presence of long wave-length perturbations is at hand.  Intuitively, one may expect that the same result does hold for every perturbed patch of the universe in the long wave-length limit. The separate universe picture states that the perturbed patches smoothed on a super-horizon scale evolve completely similar to an unperturbed universe. Therefore a complete set of equations for ${\cal N}(\mathbf{x}, t)$ and component of integrated shear ${\cal M}_{ij}(\mathbf{x}, t)$ in the near FLRW geometries are as following
\begin{align}
{\cal N}'^2&=\frac{1}{3M^{2}_{P}} {\cal E}+ \mathcal{O}(m^2,\epsilon^2) \,,
\label{12}\\
{\cal N}''+{\cal N}'^2 &=-\frac{1}{6M^{2}_{P}}({\cal E}+{\cal S})+\mathcal{O}(m^2,\epsilon^2)\,,
\label{222}\\
{\cal E}'+3{\cal N}'({\cal E}+\dfrac{1}{3}{\cal S})&= \mathcal{O}(m^2,\epsilon^2) \, ,\label{33}\\
{\cal M}''_{\chi}+3{\cal N}'{\cal M}'_{\chi}&=\dfrac{1}{M^{2}_{P}}\hat{\cal S}_{\chi} +\mathcal{O}(m^2,\epsilon^2) \, ,~~~~~~~~~~ \chi = \Vert,a,\lambda \, .
\label{M_CHI}
\end{align}

Using the $\delta M$ formula, one can find tensor perturbations, for example, through solving the above equations with appropriate initial and final conditions. In the following section the prescription for finding tensor perturbations is addressed in detail. 
\section{Calculating some observable perturbations}\label{deltaNM}
In this section, we discuss about the solutions of long wavelength perturbations. Their governing equations can be found from the perturbed equations discussed in Appendix A. These solutions help us to calculate some observable perturbations, as the powerful $\delta N$ formalism paves the way of studying a scalar gauge invariant perturbation in FRLW spacetime. Although, the $\delta N$ formalism has been extensively worked on, in the following, we review the important points behind this influential formula. Then we proceed to construct a similar formalism to calculate tensor gauge invariant perturbations and denote it $\delta M$ formalism.
\subsection{$\delta N$ formalism}  
As a convenient scalar quantity, we start with a geometric quantity, the spatial curvature perturbation, which is customary to use when super-horizon scalar perturbations are discussed. We review two different methods, discussed in literature, which relate the curvature perturbation to the number of $e$- folds computed between two time slices. First, we pursue the method adopted by Sasaki et al. in [11],  where the perturbation equations of motion are investigated in the gauge specified by $\dot{\psi}=\beta=0$. In this gauge, the dynamics of scalar field perturbation will resemble closely to that of the background, if the time coordinate is taken to be the $e$-folding number $N$ of the cosmic expansion. It can be easily seen that , in above gauge, the scalar perturbation $h_{||}$ is related to the curvature perturbation, ${\cal R}$ and has since different components of ${\cal M}_{ij}$ satisfy (\ref{M_CHI}), it has the following dynamics
\begin{equation}
\ddot{h}_{||}+3\dot{N}\dot{h}_{||}=\delta{\cal S}_{||}+{\cal O}\left(\epsilon^2, m^2\right). 
\end{equation}
Hence, up to ${\cal O}\left(\epsilon^2, m^2\right)$ and with initial condition $h_{||}\left(t_0\right)=0$, we obtain 
\begin{equation}\label{h|| solution}
h_{||}\left(t,\textbf{x}\right)=c_{||}\left(\textbf{x}\right)\int^{t}_{t_0}e^{-3N}dt+\int^{t}_{t_0}e^{-3N}\left[\int^{t}_{t_0}\delta S_{||}e^{3N}d\acute{\acute{t}}\right]d\acute{t}.
\end{equation}
In the language of [11], in which the anisotropic stress tensor is zero and $e$-folding number plays the role of time, the curvature perturbation is given by
\begin{equation}
{\cal R}\left(N,\textbf{x}\right)\vert_{\dot{\psi}=0}=c\left(\textbf{x}\right)\int^{N}_{N_0}\frac{e^{-3\acute{N}}d\acute{N}}{H}.
\end{equation}

 Applying a gauge transformation (which is a residual gauge degree of freedom corresponding to an infinitesimal time translation) given by an appropriate $\delta N$ to the perturbation in $\dot{\psi}=\beta=0$ gauge, moves ${\cal R}\left(N\right)$ to comoving curvature perturbation, ${\cal R}_c={\cal R}\left(N\right)+\delta N$. The amplitude of spatial curvature perturbation on the comoving hypersurface is then given by 
\begin{equation}
{\cal R}_c = \Delta N \vert_{\mathrm{flat} \rightarrow \mathrm{comoving}}+c\left(\textbf{x}\right)\int^{N}_{N_0}\frac{e^{-3N}}{H}d\acute{N}.
\end{equation}
This has a correction in comparison to the standard $\delta N$ formula. This correction term clearly decays in inflationary models.\\

 The authors in [11] showed that the amplitude of spatial curvature perturbation on comoving hyper-surface can be calculated from the knowledge of the background solutions, alone. In comoving gauge, the uniform $N$ slices are orthogonal to the fluid 4-velocity, $Z_N$, of a fluid element $Z$. Similarly, for a general field $Z$, with background value $\bar{Z}$, this can be described as ${\cal L}_X{Z}.Z_{N}=0$ and in case of linear perturbations it gives
\begin{equation}
\delta Z.\bar{Z}_N+\delta N \left(\bar{Z}_N\right)^2=0\Rightarrow \delta N=-\frac{\delta Z.\bar{Z}_N}{\left(\bar{Z}_N\right)^2}.
\end{equation}
Here $X$ is the comoving vector field given by $X=\frac{d}{dN}$ and ${\cal L}_X$ is the Lie derivative with respect to $X$. On the other hand, $\delta Z$ as a perturbation of the background field $\bar{Z}$, can also be constructed from the background solutions; $\delta Z_{\left({\alpha}\right)}=\partial \bar{Z}/\partial \lambda^{\alpha}$, where $\lambda^{\alpha}=\left(N,\lambda^{a}\right)$ and $\lambda^{a}$s are integration constants which distinguish various phase space solutions. The authors also showed that with slow rolling assumption, not only $c=0$ but also ${\delta Z}_{\left(a\right)}.\bar{Z}_N=0$, and the calculation result would be the famous $\delta N$ formula.\\

The second approach in calculating the curvature perturbation is given in \cite{Sugiyama:2012tj}, where the separate universe picture is valid in flat gauge specified by $\psi=h_{||}=h_{a}=0$ with $\beta=O\left(\epsilon\right)$ and $\alpha=1$. They work with gauge invariant curvature perturbation in the uniform density gauge, $\zeta\equiv\psi\vert_{\delta\rho=h_{||}=h_{a}=0}$. A gauge transformation (of a time translation type), like the one we ended up in Appendix \ref{nonlinear}, moves $\psi$ from a flat time-slice to a uniform density time-slice. In this gauge transformation, $L^i=O\left(\epsilon\right)$, where $L^i$s are the spatial degrees of freedom. This condition preserves the order of perturbation $\beta$ in gradient perturbation scheme. Equation (\ref{Npsi}) shows the $\psi$ changes under the gauge transformation in which $h_{||}=h_{a}=0$. Moving from flat gauge, $\left(\psi=h_{||}=h_{a}=0\right)$, to uniform density gauge, $\left(\delta\rho=h_{||}=h_{a}=0\right)$, gives $\zeta$ as a perturbation to the number of $e$-folds, $N$; 
\begin{equation}
\zeta\equiv\psi\vert_{\delta\rho=h_{||}=h_{a}=0}=\delta N.
\end{equation}
The authors in \cite{Sugiyama:2012tj} showed that super-horizon $\zeta$ perturbations are conserved in a universe dominated by a single scalar field, provided that either the slow roll conditions are satisfied or we neglect a decaying mode solution.\\

In zero-shear limit, the dynamics of scalar perturbations on FLRW and Bianchi spacetimes are the same{\footnote{Except that $h_{||}$ is not a pure gauge (as in FLRW) and this affects on its dynamics. This does not worry us, because it is chosen to be zero in both initial and final gauges.}}. In other words, up to $O\left(\epsilon^2, m^2\right)$, the spatial curvature perturbations on a uniform $N$ Bianchi hyper-surface are practically indistinguishable from  the one discussed above; so in Bianchi spacetimes also $\psi\vert_{\delta\rho=h_{||}=h_{a}=0}$ can be calculated by $\delta N$ formula, with the same considerations for the accuracy of the formula. $\delta N$ formalism in anisotropic inflation has been thoroughly studied in \cite{Abolhasani:2013zya}. \\ 
\subsection{$\delta M$ formalism} 
In anisotropic inflationary models, super-horizon perturbations are influenced by an anisotropic background. The decomposition of the perturbations showed that $\delta{\cal S}_{\lambda}$, plays role in the evolution of tensor perturbations. The behavior of some spatial three-tensor perturbed quantities under the infinitesimal coordinate transformation generated by the Lie derivative satisfying $L^i=O\left(\epsilon\right)$, is discussed in Appendix \ref{nonlinear}. There are some physical quantities, for example ${\cal S}_{ij}$, $\gamma_{ij}$ or $\sigma_{ij}$, which are not scalars but transform like scalars. As we show in Appendix \ref{nonlinear}, ${\cal M}_{ij}$ is one of these scalar-like objects. It is easy to see that under the above transformation, the energy density $\rho$ and $e$-folding number, ${\cal N}$ are also four-scalars.  One can use the Stewart-Walker lemma \cite{Stewart:1974uz,Sopuerta:2003rg} to show that different components of ${\cal M}_{ij}$ are gauge invariant perturbations to the FLRW  limit spacetimes under {\it general} gauge transformations. \\ 

One can construct a gauge-invariant combination of the tensor modes. In linear order, we have
\begin{align}
H_{\lambda}\equiv h_{\lambda}\vert_{\delta\rho=h_{||}=h_{a}=0}=h_{\lambda}-\frac{\sigma_{\lambda}}{H}\frac{\delta\rho}{\dot{\rho}} 
\end{align}
as a gauge invariant tensor mode \cite{Pereira:2007yy}. This is the tensor perturbation in the "uniform density gauge".  We study the conservation of this observable, in long wavelength limit, and show that unless the slow-roll conditions are violated; so it can be used to express the contribution of tensor modes in temperature and polarization anisotropies. By neglecting the $O\left(\epsilon^2, m^2\right)$ terms, the equations governing $h_{\lambda}\vert_{\delta\rho=h_{||}=h_{a}=0}$ can be obtained from (\ref{dyn.eq4.hat}) as
\begin{equation*}
\ddot{H_{\lambda}}+3\dot{N}\dot{H_{\lambda}}+3\dot{\psi}M_{\lambda}=\delta S_{\lambda}.
\end{equation*}
Since $\dot{\psi}\vert_{\delta\rho=0}=\dot{\zeta}\approx0$, the third term vanishes{\footnote {In fact, $\dot{\zeta}$ is of the order of accuracy of the $\delta N$ formalism}} and a straightforward solution is given by
\begin{equation*}
H_{\lambda}\left(t,\textbf{x}\right)=H_{\lambda}^{\left(1\right)}\left(\textbf{x}\right)+H_{\lambda}^{\left(2\right)}\left(\textbf{x}\right)\int^{t}_{t_0}e^{-3N}d\acute{t}
+\int^{t}_{t_0}e^{-3N}\left[\int^{t}_{t_0}\delta S_{\lambda}e^{3N}d\acute{\acute{t}}\right]d\acute{t},
\end{equation*}
where $H_{\lambda}^{\left(1\right)}=H_{\lambda}\left(t=t_0\right)$ and $H_{\lambda}^{\left(2\right)}=\left(\dot{H_{\lambda}}e^{3N}\right)_{t=t_0}$. Obviously, the solution proportional to $H_{\lambda}^{\left(2\right)}$ is an adiabatic decaying solution in inflationary models. The third term in the right hand side would be a decaying mode, if $\int^{t}_{t_0}\delta S_{\lambda}e^{3N}d\acute{t}$ evolves slower than $e^{3N}$. Thus, $H_{\lambda}\equiv h_{\lambda}\vert_{\delta\rho=h_{||}=h_{a}=0}$ would be conserved, if we neglect the decaying mode solutions. In the following, we give a recipe for calculating this observable. It goes without saying that like her twin formula, $\delta N$, the validity of the result depends on the model. This is naturally expected to be valid in any inflationary scenario, in which the amplitude of variation of anisotropic stress is of order ${\cal O}\left(\epsilon^1 \mathrm {or} \epsilon^2 \right)$\footnote{In other words, $\delta S_{\lambda}$ should be a super-horizon mode in a inflationary background}.\\

The relation of different components of $h_{ij}$ with the amount of $M_{ij}$ changes under the gauge transformation generated by $t_i\rightarrow t_f=t_i+T$ and $L^i=O\left(\epsilon\right)$ can be read from equation (\ref{Mh}).
  It is easy to see that for variations of the background $M_{\lambda}$ between an initial flat hypersurface in which $\psi=h_{||}=h_{a}=0$  and a final uniform density hyper-surface described by
\begin{eqnarray}{\cal M}_{\lambda}\left(t_i,\mathbf{x}\right)=M_{\lambda}\left(t_i\right)+h_\lambda\left(t_i,\mathbf{x}\right)\rightarrow M_{\lambda}\left(t_f\right)+{\cal O}(\epsilon^2,m^2),\\
\end{eqnarray}
the tensor perturbation in comoving gauge are given by
\begin{align}\label{delta M}
H_{\lambda}(t_f,\mathbf{x})= \delta M_{\lambda}(t_i,t_f)
\end{align}
 Equation (\ref{delta M}) is $\delta M$ formula appropriate for calculating long wavelength tensor perturbations. This is a geometric, gauge invariant identity relating the tensor perturbations in Bianchi spacetime to the background.\\

 $H_{\lambda}$ can be readily related to the perturbed initial fields in the initial gauge. In a gauge transformation from the flat to the comoving gauge, the energy density transform as ${\rho}\left(t_i,\mathbf{x}\right)\rightarrow\bar{\rho}\left(t_f\right)$.  We recall that, $L^i=0$; so there is no ambiguity with respect to the spatial gauge degrees of freedom. In separate universe picture, the perturbed quantities are related to their unperturbed values, using the initial fields solutions. Therefore, for the perturbed energy density we have
\begin{align}\label{rho SU}
{\rho}\left(t_i,\mathbf{x}\right)=\bar{\rho}\left(t_i,\Phi_I\left(t_0,\mathbf{x}\right)\right).
\end{align}
For the sake of brevity, we used $\Phi_I$ to refer to the values of scalar, vector or higher order fields, which take part in the problem as well as their derivatives. Here, $t_0$ refers to the initial time coordinate at the time all the $\Phi_I$ fields were sufficiently outside their horizon. As it is shown in (\ref{N}), energy density is a scalar and we have $\rho\left(t_i,\mathbf{x}\right)=\bar{ \rho}\left(t_f\right)$. This added to (\ref{rho SU}) can be used to yield
\begin{align}\label{tf vs ti}
t_i=t_f\left(\bar{\rho},\Phi_I\left(t_0,\mathbf{x}\right)\right).
\end{align}
Equation (\ref{tf vs ti}) shows that i) the final hyper-surface coincides with uniform density and ii) the "`functional form"' of the time coordinates in initial and final hypersurface are the same. With these results, equation (\ref{delta M}) is given by
\begin{equation}
H_{\lambda}=M_{\lambda}\left(\bar{\rho},\Phi_I\left(t_0,\mathbf{x}\right)\right)-M_{\lambda}\left(\bar{\rho},\bar{\Phi}_I\left(t_0\right)\right)
\end{equation}
It may be written in terms of the perturbations of initial fields, $\delta \Phi_I\left(t_0,\mathbf{x}\right)\equiv\Phi_I\left(t_0,\mathbf{x}\right)-\bar{\Phi}_I\left(t_0\right)$,
\begin{align}\label{delta M final}
H_{\lambda}=M_{\lambda,I}\delta \Phi_I\left(\mathbf{x}\right)+\frac{1}{2}M_{\lambda,{IJ}}\delta {\Phi}_I\left(\mathbf{x}\right)\delta \Phi_J\left(\mathbf{x}\right)+\ldots,
\end{align}
where 
\begin {equation}
M_{\lambda,I}\equiv\frac{\partial M_{\lambda}\left[\bar{\rho},\bar{\Phi}_J\right]}{\partial\bar{\Phi}_I},\ \ \ \
M_{\lambda,{IJ}}\equiv\frac{\partial^2 M_{\lambda}\left[\bar{\rho},\bar{\Phi}_K\right]}{\partial\bar{\Phi}_I\partial\bar{\Phi}_J},
\end{equation}
and  ellipsis in (\ref{delta M final}) indicates the higher order terms in Taylor expansion of background function. $\delta M$ formalism in this form relates $H$ to the perturbations of the initial fields. It is important to note that all the perturbations are computed in the flat gauge.  $\zeta$ is also related to the initial scalar field perturbations computed in flat gauge \cite{Sugiyama:2012tj},
\begin{equation}
\zeta=N_{I}\delta \Phi_I\left(\mathbf{x}\right)+\frac{1}{2}N_{{IJ}}\delta {\Phi}_I\left(\mathbf{x}\right)\delta \Phi_J\left(\mathbf{x}\right)+\ldots,
\end{equation}
where
\begin {equation}
N_{I}\equiv\frac{\partial N\left[\bar{\rho},\bar{\Phi}_J\right]}{\partial\bar{\Phi}_I},\ \ \ \
N_{{IJ}}\equiv\frac{\partial^2 N\left[\bar{\rho},\bar{\Phi}_K\right]}{\partial\bar{\Phi}_I\partial\bar{\Phi}_J}.
\end{equation}
As stated above, $\delta M$ formalism holds under the assumption that the so called adiabatic decaying modes can be neglected. One can recognize that in models with negligible ansiotropic stress $\hat{\cal S}_{\lambda}$, the variation $\delta M_{\lambda}$ is of the order of the accuracy of the formalism. However, we are particularly interested in the effect of non vanishing ansiotropic stress on tensor perturbations, which in principle could significantly change total tensor perturbation in anisotropic inflation models. We showed that non-adiabatic tensor perturbations are found by only knowing the background ${M_\lambda}$ solution just like the comoving curvature perturbation is found by knowing the total background $e$-folding number.  

\section{Summary and Conclusion}
\label{discussions}
We have applied the gradient expansion to the evolution equations of perturbations in an anisotropic (Bianchi-{\it I}) universe. As a consequence, we came up with a consistent separate universe approach to the perturbation theory. This has been demonstrated by the form invariance of Einstein constraint and dynamical equations at long-wavelength perturbation limit.

In particular, we have found how the different classes of perturbations can be absorbed in homogeneous but anisotropic background parameters. To be specific, the background geometrical shear can be redefined in such a way that absorb long-wavelength vector and tensor perturbations comparable to what the scale factor does for scalar perturbation in the standard $\delta N$ formalism. 

We have obtained a powerful tool called $\delta M$ formalism which relates the amplitude of spatial perturbations to the change in $M_{ij}$, which in turn is defined to measure the homogeneous changes in Euclidean metric. In particular, the change in the amplitude of a gravitational wave, with comoving wavelength $k^{-1}$, going from one time slice to another is related to variations of integrated shear $M_{\lambda}=\int_{t_0}^{t} \sigma_{\lambda} \, \mathrm{d}t'$ calculated between two initial flat and final uniform density hyper-surfaces. It should be mentioned that in contrast to its analogous $\delta N$ formula, this relation is restricted to {\it linear} perturbations on FLRW universe. Even though we started the analysis non-perturbatively, we neglected $\sigma^2$ terms in the right hand side of equations (\ref{sigma_Vert}-\ref{sigma_lambda}); this will inevitably result in a perturbative formula, which is restricted to weak shear limit. 

Apart from the theoretical interest in studying long-wavelength perturbations in the anisotropic background, the proposed $\delta M$ formalism is a powerful method for studying the perturbations in anisotropic inflationary models \cite{Aniso:Inflation}. Specifically, it is a tool for dealing non-trivial interaction between scalar and tensor modes showing up in anisotropic inflationary models. In a word, employing the $\delta M$ formalism substantially simplifies the calculations of the correlation functions of the linear perturbations in these models. We believe that inclusion of first order anisotropy corrections to FLRW model is accurate enough for the analysis of the observational features of an anisotropic model on the CMB. The likelihood of detecting anisotropies with higher accuracy motivates the development of different methods of computing the perturbations.

\acknowledgments
We acknowledge fruitful discussions with Misao Sasaki and Hassan Firouzjahi. We would like to thank Mohammad Nouri-zonoz, Kamal Hajian for precious comments. A.T.A thanks ``Halghe" cosmology group for stimulating discussions. We wish to thank school of physics of IPM for hospitality. A.T.A and N.A thank university of Tehran for supporting this project under the grants provided by the research council. A.A.A is partially supported by deputy of research of the SUT.

\appendix

\section{Conformal 3+1 Decomposition}
\label{conf3+1}
In the ADM decomposition \cite{Arnowitt:1962hi,Gourgoulhon_book}, the metric is expressed as \eqref{ADMmetric}, with the following matrix form
\begin{equation*}
g_{\alpha \beta}= \begin{bmatrix}
-\alpha^2+\beta^{k}\beta_{k} &  & \beta_{i}  \\
\\
\beta_{j} & & \gamma_{ij}   \\
\end{bmatrix} \, , ~~~~~~ \quad
g^{\alpha \beta}= \begin{bmatrix}
\dfrac{-1}{\alpha^2} & \dfrac{\beta^{i}}{\alpha^2}  \\
\\
\dfrac{\beta^{j}}{\alpha^2} & ~~~~\gamma^{ij}- \dfrac{\beta^{i}\beta^{j}}{\alpha^2} \\
\end{bmatrix} \, .
\end{equation*}

Any smooth physical space-time in the standard ADM formalism can be decomposed into a unit time-like vector $\boldsymbol{n}$ and a constant time hyper-surfaces, $\boldsymbol{\Sigma}_{t}$, which is normal to it. The components of $\boldsymbol{n}$ in the space-time \eqref{ADMmetric} are given by 
\begin{align}
n_{\mu}=[-\alpha\,,0] \, , \quad n^{\mu}=\dfrac{1}{\alpha}\left[1,-\beta^{i} \right] \, .
\label{n}
\end{align}
The induced metric (the first fundamental form) on $\boldsymbol{\Sigma}_{t}$, is defined as
\begin{align}
\Sigma_{\mu\nu}=g_{\mu\nu}+n_{\mu}n_{\nu} \, .
\end{align}
Therefore the spatial part of induced metric is $\gamma_{ij}$. The extrinsic curvature (the second fundamental form) of $\boldsymbol{\Sigma}_{t}$, $\boldsymbol{\mathcal{K}}$ is defined as
\begin{align}
\mathcal{K}_{ij} \equiv -\nabla_{i}n_{j}=\frac{1}{2\alpha}(D_{i}\beta_{j}+D_{j}\beta_{i}-\dot{\gamma}_{ij}) \, ,
\label{extrinsic1}
\end{align}
where the dots denote time derivatives with respect to cosmic time $t$, and $\boldsymbol{\nabla}$ and $\boldsymbol{D}$ are the covariant differential operators constructed by using $g_{\mu\nu}$ and $\gamma_{ij}$ by means of related Christoffel symbols $\Gamma^{\alpha}_{\mu \nu}$ and ${}^{3}\Gamma^{k}_{ij}$, respectively. One can check that $\mathcal{K}_{ij}=\alpha \Gamma^{0}_{ij}$.

In the context of (3+1)-decomposition of Einstein equations, the dynamical variables are the spatial three metric, $\gamma_{ij}$, and the extrinsic curvature, $\mathcal{K}_{ij}$. The components of $(n,n) \equiv G_{\mu \nu} n^{\mu}n^{\nu}$, and $(n,i) \equiv G_{\mu \nu} n^{\mu} \Sigma^{\nu}_{i}$, of the Einstein equations yield Hamiltonian and momentum constraints, respectively and $(i,j)\equiv G_{\mu \nu} \Sigma^{\mu}_{i} \Sigma^{\nu}_{j}=G_{ij}$ components give the evolution equations for extrinsic curvature. The evolution equations for the spatial part of induced metric are given by the definitions of the extrinsic curvature \eqref{extrinsic1}.

To write down Einstein equations, $G_{\mu\nu}=M^{-2}_{P}T_{\mu\nu}$ where $ M^{-2}_{P} = 8 \pi G /c^4$ is the reduced Planck mass and $G$ is the gravitational constant, we need an energy momentum tensor as a source. It is convenient to represent the independent components of the energy momentum tensor as energy density $\mathcal{E}=T_{\mu \nu} n^{\mu}n^{\nu}$, momentum density $\mathcal{J}_{i} = T_{\mu \nu} n^{\mu} \Sigma^{\nu}_{i}$, and stress tensor $\mathcal{S}_{ij}=T_{\mu \nu} \Sigma^{\mu}_{i} \Sigma^{\nu}_{j}$.

In the ADM decomposition the four constraints (one Hamiltonian and three momentum constraints) are
\begin{align}
{}^{3}\mathcal{R}+\mathcal{K}^2-\mathcal{K}_{ij}\mathcal{K}^{ij}&=\frac{2}{M^{2}_{P}} \mathcal{E} \, ,
\label{const.eq1}\\
D_{j}\mathcal{K}^{j}_{i}-D_{i}\mathcal{K}&=-\frac{1}{M^{2}_{P}}\mathcal{J}_{i} \, .
\label{const.eq2}
\end{align}
Moreover, two sets of dynamical equations for $\gamma_{ij}$ and $\mathcal{K}_{ij}$ can be found as follows
\begin{align}
\partial_{t}\gamma_{ij}&=-2\alpha\mathcal{K}_{ij}+D_{i}\beta_{j}+D_{j}\beta_{i} \, ,\label{dyn.eq1}\\
(\partial_{t}-\beta^{k}\partial_{k})\mathcal{K}_{ij}&=\mathcal{K}_{kj}\partial_{i}\beta^{k}+\mathcal{K}_{ik}\partial_{j}\beta^{k}
-D_{i}D_{j}\alpha+\alpha({}^{3}\mathcal{R}_{ij}+\mathcal{K}\,\mathcal{K}_{ij}-2\mathcal{K}_{ik}\mathcal{K}^{k}_{j})
\nonumber\\
&~~~~~~~~~~~~
~~~~~~~~~~~~~
~~~~~~~~~~~~
~~~-\frac{\alpha}{M^{2}_{P}}(\mathcal{S}_{ij}-\frac{1}{3}\gamma_{ij}\gamma^{kl}\mathcal{S}_{kl}) \, ,\label{dyn.eq2}
\end{align}
where ${}^{3}\mathcal{R}_{ij}$ and ${}^{3}\mathcal{R}$ are Ricci tensor and Ricci scalar of spatial three metric respectively and $\mathcal{K}$ is a trace part of $\mathcal{K}_{ij}$. The indices of $\mathcal{K}_{ij}$ are raised by $\gamma^{ij}$.

York \cite{YORK} has shown that the two degrees of freedom of the gravitational field are carried by the conformal equivalence classes \footnote{All metrics that can be related to $\boldsymbol{\gamma}$ by a conformal transformation such as \eqref{Conf_trans} are in the same class} of spatial three metrics. Since we are going to focus on tensor perturbations, the metric must be further decomposed, We follow Lichnerowicz \cite{Gourgoulhon_book} who has decomposed the spatial metric $\boldsymbol{\gamma}$ into a positive scalar field $\boldsymbol{\Theta}$ and an auxiliary metric $\boldsymbol{\tilde{\gamma}}$:
\begin{align}
\boldsymbol{\gamma}=e^{2\boldsymbol{\Theta}} ~ \boldsymbol{\tilde{\gamma}} \, .
\label{Conf_trans}
\end{align}
This relation is a conformal transformation and $\boldsymbol{\tilde{\gamma}}$ a conformal spatial three metric. We go further through enfolding determinant part of $\boldsymbol{\gamma}$ into $e^{2\mathcal{N}}$ and make an unit-determinant conformal metric $\boldsymbol{\hat{\gamma}}$, $\hat{\gamma}=\det{[\hat{\gamma}_{ij}]}=1$,
\begin{align}
\gamma_{ij}=e^{2\mathcal{N}} \hat{\gamma}_{ij} \, .
\end{align}
So we have $\hat{\gamma}_{ij}=\gamma^{-1/3} ~ \gamma_{ij}$, where $\gamma=\det{[\gamma_{ij}]}$. In fact $\boldsymbol{\hat{\gamma}}$ is a tensor density of weight -2/3. This unit-determinant metric can be used to decompose the spatial metric into the expansion and anisotropic parts as Misner and Hawking have done for studying Bianchi space-times \cite{Misner:1967uu,Hawking:1968zw}.

For further simplifications of equations, all the spatial quantities $(\mathcal{K}_{ij}, {}^{3}\mathcal{R}_{ij}, \mathcal{S}_{ij})$  are decomposed into trace and traceless parts
\begin{align}
\mathcal{K}_{ij}&=\frac{1}{3}\mathcal{K}~\gamma_{ij}+e^{2\mathcal{N}}\hat{\mathcal{A}}_{ij} \, ,
\label{K_decomposed}\\
{}^{3}\mathcal{R}_{ij}&=\frac{1}{3}{}^{3}\mathcal{R} \gamma_{ij}+e^{2\mathcal{N}}\,\,{}^{3}\hat{\mathcal{R}}_{ij} \, ,\\
\mathcal{S}_{ij}&=\frac{1}{3}\mathcal{S}~\gamma_{ij}+e^{2\mathcal{N}}\hat{\mathcal{S}}_{ij} \, ,
\label{stress2}
\end{align}
where $\mathcal{K}=\gamma^{ij}\mathcal{K}_{ij}$, ${}^{3}\mathcal{R}=\gamma^{ij}{}^{3}\mathcal{R}_{ij}$ and $\mathcal{S}=\gamma^{ij}\mathcal{S}_{ij}$.
The indices of the traceless quantities, $(\hat{\mathcal{A}}_{ij}, {}^{3}\hat{\mathcal{R}}_{ij}, \hat{\mathcal{S}}_{ij})$ are raised and lowered by $\hat{\gamma}^{ij}$ and $\hat{\gamma}_{ij}$, respectively. The explicit forms of $\mathcal{K}$ and $\hat{\mathcal{A}}_{ij}$ are given by
\begin{align}
\mathcal{K}&=\mathcal{K}^{i}_{i}=\gamma^{ij}\mathcal{K}_{ij}=\dfrac{1}{\alpha}(D_{k}\beta^{k}-3\dot{\mathcal{N}}) \, ,
\label{K}\\
\hat{\mathcal{A}}_{ij}&=e^{-2\mathcal{N}}(\mathcal{K}_{ij}-\frac{1}{3}\mathcal{K}\gamma_{ij})=-\frac{1}{2\alpha} \dot{\hat{\gamma}}_{ij}+\dfrac{e^{-2\mathcal{N}}}{2\alpha}(D_{i}\beta_{j}+D_{j}\beta_{i}-\dfrac{2}{3}D_{k}\beta^{k}\gamma_{ij}) \, .
\label{A}
\end{align}
It is easy to show that $\hat{\mathcal{A}}_{ij}$ is traceless,
\begin{align}
\hat{\mathcal{A}}^{i}_{~i}=-\frac{1}{2\alpha}\hat{\gamma}^{ij} \dot{\hat{\gamma}}_{ij}=-\frac{1}{2\alpha}\dot{\hat{\gamma}}=0 \, .
\label{unitdet.tracefree}
\end{align}
${}^{3}\hat{\mathcal{R}}_{ij}$,
$\mathcal{S}$ and $\hat{\mathcal{S}}_{ij}$ can be interpreted as anisotropic Ricci tensor,  isotropic and anisotropic pressure, respectively.

By applying the above decompositions to Eqs. \eqref{const.eq1}-(\ref{dyn.eq2}), one finds the equations governing the newly defined fields. The constraint equations turn into
\begin{align}
{}^{3}\mathcal{R}+\frac{2}{3}\mathcal{K}^2-\hat{\mathcal{A}}_{ij}\hat{\mathcal{A}}^{ij}&=\frac{2}{M^{2}_{P}}\mathcal{E} \, ,
\label{const.eq1.hat}\\
D_{j}\hat{\mathcal{A}}^{j}_{i}-\frac{2}{3}\partial_{i}\mathcal{K}&=-\frac{1}{M^{2}_{P}}\mathcal{J}_{i} \, .
\label{const.eq2.hat}
\end{align}
The dynamical equations for the spatial metric $(\mathcal{N},\hat{\gamma}_{ij})$ are
\begin{align}
(\partial_{t}-\beta^{k}\partial_{k})\mathcal{N} &= \frac{1}{3}(-\alpha \mathcal{K}+\partial_{k}\beta^{k}) \, ,
\label{dyn.eq1.hat}\\
(\partial_{t}-\beta^{k}\partial_{k})\hat{\gamma}_{ij}&=-2\alpha\hat{\mathcal{A}}_{ij}+\hat{\gamma}_{ik}\partial_{j}\beta^{k}
+\hat{\gamma}_{jk}\partial_{i}\beta^{k}-\frac{2}{3}\hat{\gamma}_{ij}\partial_{k}\beta^{k} \, .
\label{dyn.eq3.hat}
\end{align}
The dynamical equations for extrinsic curvature $(\mathcal{K},\hat{\mathcal{A}}_{ij})$ are
\begin{align}
(\partial_{t}-\beta^{k}\partial_{k})\mathcal{K} &=\alpha(\hat{\mathcal{A}}_{ij}\hat{\mathcal{A}}^{ij}+\frac{1}{3}\mathcal{K}^2)-\gamma^{ij}D_{i}D_{j}\alpha+\frac{\alpha}{2M^{2}_{P}}(\mathcal{E}+\mathcal{S}) \, ,
\label{dyn.eq2.hat}\\
(\partial_{t}-\beta^{k}\partial_{k})\hat{\mathcal{A}}_{ij}&=\alpha(\mathcal{K}\hat{\mathcal{A}}_{ij}-2\hat{\mathcal{A}}_{ik}\hat{\mathcal{A}}^{k}_{j})-\frac{\alpha}{M^2_{P}} \hat{\mathcal{S}}_{ij}+\alpha\,\, {}^{3}\hat{\mathcal{R}}_{ij}-e^{-2\mathcal{N}}(D_{i}D_{j}\alpha-\frac{1}{3}\gamma_{ij}D_{k}D^{k}\alpha)\nonumber\\
&~~~~~+\hat{\mathcal{A}}_{ik}\partial_{j}\beta^{k}
+\hat{\mathcal{A}}_{jk}\partial_{i}\beta^{k}-\frac{2}{3}\hat{\mathcal{A}}_{ij}\partial_{k}\beta^{k} \, .
\label{dyn.eq4.hat}
\end{align}

The above equations in addition to matter equations constitute a complete set to solve. Matter obeys the field equations and the continuity equations. The continuity equations, $\nabla_{\nu}T^{\nu\mu}=0$, are given by
\begin{align}
(\partial_{t}-\beta^{k}\partial_{k})\mathcal{E} &- \alpha \Big(\mathcal{K}(\mathcal{E}+\dfrac{1}{3}\mathcal{S})+\hat{\mathcal{K}}_{ij}\hat{\mathcal{S}}^{ij}\Big) +\alpha D_{i}\mathcal{J}^{i}+2\mathcal{J}^{i}D_{i}\alpha=0 \, ,
\label{continuity_E}\\
(\partial_{t}-\beta^{k}\partial_{k})\mathcal{J}_{i} &- \alpha \Big(\mathcal{K}\mathcal{J}_{i}-D_{j}\mathcal{S}_{i}^{j}\Big) +\Big( \mathcal{E} \delta^{j}_{i}+\mathcal{S}^{j}_{i}\Big) D_{j}\alpha=0 \, .
\label{continuity_J}
\end{align}

The symmetric unit-determinant conformal metric $\boldsymbol{\hat{\gamma}}$ is usually expressed in terms of a symmetric trace-free matrix $\boldsymbol{\mathcal{M}}$ as
\begin{align}
\hat{\gamma}_{ij}=(e^{2\boldsymbol{\mathcal{M}}})_{ij} \, .
\label{hat_gamma_M}
\end{align}
Then, as is well known $\det[\hat{\gamma}_{ij}]=e^{2 Tr \boldsymbol{\mathcal{M}}}=1$ where $Tr \boldsymbol{\mathcal{M}}=\mathcal{M}_{i}^{i}=0$. The matrix exponential is defined via Taylor expansion.
\begin{align}
(e^{2\boldsymbol{\mathcal{M}}})_{ij} &=\sum_{n=0}^{+\infty} \frac{(2\boldsymbol{\mathcal{M}})^{n}_{ij}}{n!}
\nonumber\\
&=\delta_{ij} + 2 \mathcal{M}_{ij} + \dfrac{2^2}{2!} \mathcal{M}^{k}_{i} \mathcal{M}_{kj} + \dfrac{2^3}{3!} \mathcal{M}^{k}_{i} \mathcal{M}_{kl} \mathcal{M}^{l}_{j} + ... \, ,
\label{MatricEXP}
\end{align}
in which $\mathcal{M}^{i}_{j}=\delta^{ik} \mathcal{M}_{kj}$.

\subsection{Fluid and Scalar Field Description }
\label{Fluid_Scalar}
If matter is described as a fluid with 4-velocity $u^{\mu}$, its energy-momentum tensor $T_{\mu \nu}$ will be decomposed uniquely as
\begin{align}
T_{\mu \nu}= \rho u_{\mu}u_{\nu}+pP_{\mu\nu}+q_{\mu}u_{\nu}+q_{\nu}u_{\mu}+\pi_{\mu \nu} \, ,
\label{EMT}
\end{align}
supplemented with the following conditions 
\begin{align}
q_{\mu}u^{\mu} = 0 =\pi_{\mu \nu}u^{\nu} \, ,~\pi^{\mu}{}_{\mu} = 0 \, ,~\pi_{\mu \nu}=\pi_{\nu \mu} \, ,~\pi_{\mu \nu}P^{\mu \nu} = 0 \, .
\label{matter_condition}
\end{align}
Here $\rho$ is the relativistic energy density relative to $u^{\mu}$ (measured by an observer at rest with the fluid), $p$ is the isotropic pressure, $q^\mu$ usually is referred  to as ``heat conduction'', which is also the energy flux (relativistic momentum density) relative to $u^\mu$ and $\pi_{\mu \nu}$ is the trace-free  anisotropic pressure (stress) tensor. Also $P_{\mu \nu}=g_{\mu\nu}+u_{\mu}u_{\nu}$ is the projection tensor onto the surface of perpendicular to $u_{\mu}$.

The components of fluid's four-velocity,
\begin{align}
u^{\mu}=\dfrac{\mathrm{d}x^{\mu}}{\mathrm{d}\tau},~~u^{\alpha}u_{\alpha}=-1 \, ,
\end{align}
where $\tau$ is proper time measured along the world line, are given by
\begin{align}
u^{o}&=[\alpha^{2}-(\beta^{k}+v^{k})(\beta_{k}+v_{k})]^{-1/2} \, ,\nonumber\\
u^{i}&=u^{o}v^{i} \, ,\nonumber\\
u_{o}&=-u^{o}[\alpha^{2}-\beta^{k}(\beta_{k}+v_{k})] \, ,\nonumber\\
u_{i}&=u^{o}(v_{i}+\beta_{i}) \, .
\label{4velocity}
\end{align}
Here $v^{i}$ is the spatial velocity, $v^i \equiv \mathrm{d}x^{i}/\mathrm{d}t=u^i/u^0$, and $v_{i}=\gamma_{ij}v^{j}$.

The hydrodynamic equations are the energy (continuity equation) and momentum (Euler equations) conservation which can be extracted form $\nabla_{\nu}T^{\nu\mu}=0$ as follow
\begin{align}
-u_{\mu}\nabla_{\nu}T^{\nu\mu}&=0 \, ,\\
(g^{\alpha}_{\mu}+u^{\alpha}u_{\mu})\nabla_{\nu}T^{\nu\mu}&=0 \, .
\end{align}

One can think of a scalar field instead of a fluid. If we assume that the Universe is filled with a scalar field $\varphi$, described by the following Lagrangian
\begin{equation}
\mathcal{L} = - \frac{1}{2} \partial^{\mu} \varphi \partial_{\mu} \varphi - V \, ,
\label{lagrangianfi}
\end{equation}
the stress-energy tensor of it reads as
\begin{equation}
T_{\mu\nu} =\partial_{\mu}\varphi \partial_{\nu} \varphi
+ g_{\mu\nu} \left( - \frac{1}{2} \partial^{\alpha} \varphi \partial_{\alpha} \varphi  - V\right) \, ,
\label{EMfi}
\end{equation}
and its equation of motion as
\begin{equation}
\frac{1}{\sqrt{g}} \partial_{\mu} \left(\sqrt{g} g^{\mu\nu} \partial_{\nu} \varphi \right) - V_{\varphi} = 0 \, ,
\label{EoMfi}
\end{equation}
where $V_{\varphi}=\dfrac{\mathrm{d}V}{\mathrm{d}\varphi}$ and $g=\det[g_{\mu\nu}]$.

\section{$\boldsymbol{M}$ as an Integrated Shear}
\label{communication}

In this appendix we investigate conditions under which $\boldsymbol{M}$ can be interpreted as integrated shear. The bars over the background shear and the background unit-determinant metric are omitted here for brevity.

Before moving on, we will derive the explicit form of shear in terms of matrix $M^{i}_{j}$. We start from the symmetry property of $\hat{\gamma}_{ij}$, $\hat{\gamma}^{ij}$ and $\hat{\sigma}_{ij}$,
\begin{align}
\hat{\sigma}^{k}_{i} \hat{\gamma}_{kj} = \hat{\sigma}^{k}_{j} \hat{\gamma}_{ki} \, ,
\end{align}
to show $ \hat{\sigma}^{i}_{j} = \hat{\gamma}^{ik} \hat{\sigma}_{kj} = \hat{\sigma}_{jk} \hat{\gamma}^{ki}$. With the help of $\hat{\gamma}^{ij}=(e^{-2M})^{ij}$ and applying Baker-Hausdorff formula \footnote{$\dfrac{\mathrm{d}}{\mathrm{d}t}e^{M}\,e^{-M} =\dot{M}+\dfrac{1}{2!}[M,\dot{M}]+\dfrac{1}{3!}\big[M,[M,\dot{M}]\big]+ ... \, .$}
into $\hat{\sigma}^{i}_{j} = \hat{\sigma}_{jk} \hat{\gamma}^{ki}$ we get
\begin{align}
\hat{\sigma}^{i}_{j} &= \dfrac{1}{2} \dfrac{\mathrm{d}}{\mathrm{d}t}(e^{2\boldsymbol{M}})_{jk} \,\, (e^{-2\boldsymbol{M}})^{ki} \nonumber \\ &=
\dfrac{1}{2}
\sum_{n=\text{1,3,5,...}}^{\infty}\dfrac{2^{n}}{n!} \underbrace{\bigg[\boldsymbol{M},\Big[\boldsymbol{M}\cdots[\boldsymbol{M},}_\text{n-1}\boldsymbol{\dot{M}}
\underbrace{ ]\Big]\cdots\bigg]^{i}_{j}}_\text{n-1}
+\dfrac{1}{2}
\sum_{n=\text{2,4,6,...}}^{\infty}\dfrac{2^{n}}{n!} \underbrace{\bigg[\boldsymbol{M},\Big[\boldsymbol{M}\cdots[\boldsymbol{M},}_\text{n-1}\boldsymbol{\dot{M}}
\underbrace{ ]\Big]\cdots\bigg]^{i}_{j}}_\text{n-1} \, .
\label{sigma^i_j1}
\end{align}
On the other hand $\hat{\sigma}^{i}_{j} = \hat{\gamma}^{ik} \hat{\sigma}_{kj}$ gives
\footnote{$e^{-M} \dfrac{\mathrm{d}}{\mathrm{d}t}e^{M}=\dot{M}+\dfrac{1}{2!}[\dot{M},M]+\dfrac{1}{3!}\big[[\dot{M},M],M\big]+ ... \, .$}
\begin{align}
\hat{\sigma}^{i}_{j} &= \dfrac{1}{2} (e^{-2\boldsymbol{M}})^{ik} \,\,  \dfrac{\mathrm{d}}{\mathrm{d}t}(e^{2\boldsymbol{M}})_{kj} \nonumber\\ &=
\dfrac{1}{2}
\sum_{n=\text{1,3,5,...}}^{\infty}\dfrac{2^{n}}{n!} \underbrace{\bigg[\cdots\Big[[}_\text{n-1}\boldsymbol{\dot{M}}
\underbrace{,\boldsymbol{M}]\cdots \boldsymbol{M} \Big],\boldsymbol{M}\bigg]^{i}_{j}}_\text{n-1}
+ \dfrac{1}{2}
\sum_{n=\text{2,4,6,...}}^{\infty}\dfrac{2^{n}}{n!} \underbrace{\bigg[\cdots\Big[[}_\text{n-1}\boldsymbol{\dot{M}}
\underbrace{,\boldsymbol{M}]\cdots \boldsymbol{M} \Big],\boldsymbol{M}\bigg]^{i}_{j}}_\text{n-1}\,.
\label{sigma^i_j2}
\end{align}
By using $[\boldsymbol{A},\boldsymbol{B}]=-[\boldsymbol{B},\boldsymbol{A}]$ and equating (\ref{sigma^i_j1}) and (\ref{sigma^i_j2}) we have
\begin{align}
\sum_{n=\text{2,4,6,...}}^{\infty}\dfrac{2^{n}}{n!} \underbrace{\bigg[\cdots\Big[[}_\text{n-1}\boldsymbol{\dot{M}}
\underbrace{,\boldsymbol{M}]\cdots \boldsymbol{M} \Big],\boldsymbol{M}\bigg]^{i}_{j}}_\text{n-1}=0 \, .
\label{vanish_even}
\end{align}
Therefore the shear will be
\begin{align}
\hat{\sigma}^{i}_{j} &= \dot{M}^{i}_{j}
+ \dfrac{1}{2} \sum_{n=\text{3,5,...}}^{\infty}\dfrac{2^{n}}{n!} \underbrace{\bigg[\cdots\Big[[}_\text{n-1}\boldsymbol{\dot{M}}
\underbrace{,\boldsymbol{M}]\cdots \boldsymbol{M} \Big],\boldsymbol{M}\bigg]^{i}_{j}}_\text{n-1}\,.
\label{Sigma_M}
\end{align}

In Sec. \ref{Small_shear_expansion} We defined $m$ as a measure of anisotropy. Therefore \eqref{Sigma_M} results in \footnote{We note that $\hat{\sigma}_{ij}=\hat{\gamma}_{ik}\hat{\sigma}^{k}_{j}=\dot{M}_{ij}+\mathcal{O}(m^2H)$, where $M_{ij}=M^{i}_{j}\,$.}
\begin{align}
\hat{\sigma}^{i}_{j}=\dot{M}^{i}_{j}+\mathcal{O}(m^3H)\, .
\label{shear}
\end{align}
In general, $M^{i}_{j}$ cannot be considered as an integrated shear. However, if $[\boldsymbol{M},\dot{\boldsymbol{M}}]=0$, this would be the case. There are some trivial cases that $[\boldsymbol{M},\dot{\boldsymbol{M}}]=0$, for example: $\boldsymbol{M}$ is a diagonal matrix or when time independence of $\boldsymbol{M}$ can be factored out, e.g. $M^{i}_{j} = e^{i \omega t} G^{i}_{j}$, where $G^{i}_{j}$ is a constant matrix.

Let us investigate some other situations in which $[\dot{\boldsymbol{M}},\boldsymbol{M}]=0$, $\big[[\dot{\boldsymbol{M}},\boldsymbol{M}],\boldsymbol{M}\big]=0$, ... . A trace-less symmetric matrix $\boldsymbol{M}$ may be expressed in terms of an orthogonal matrix $\boldsymbol{O}$ and a diagonal trace-free matrix $\boldsymbol{D}$,
\begin{align}
\boldsymbol{M}=\boldsymbol{O}\boldsymbol{D}\boldsymbol{O}^{T} \, .
\end{align}
Then
\begin{align}
\dot{\boldsymbol{M}}=\boldsymbol{O}[\boldsymbol{\Omega}\boldsymbol{D} + \dot{\boldsymbol{D}}-\boldsymbol{D}\boldsymbol{\Omega}]\boldsymbol{O}^{T} \, ,
\end{align}
where antisymmetric $\boldsymbol{\Omega}=\boldsymbol{O}^{T}\dot{\boldsymbol{O}}$, represents the angular velocity of the principal axes of $\boldsymbol{M}$. Demanding
\begin{align}
[\dot{\boldsymbol{M}},\boldsymbol{M}]&=0 \, ,\\ \big[[\dot{\boldsymbol{M}},\boldsymbol{M}],\boldsymbol{M}\big]&=0 \, ,\\
&\vdots\nonumber
\end{align}
results in
\begin{align}
\Omega_{ij}(d_{(i)}-d_{(j)})^2&=0 \, ,
\label{omega_d_1}\\
\Omega_{ij}(d_{(i)}-d_{(j)})^3&=0 \, ,\label{omega_d_2}\\
&\vdots\nonumber
\end{align}
where $D_{ij}=d_{(i)}\delta_{ij}$, and there is no summation on $(i)$ and $(j)$. Therefore, requiring that $\Big[...\big[[\dot{\boldsymbol{M}},\boldsymbol{M}],\boldsymbol{M}\big],...\Big]=0$ results in $[\dot{\boldsymbol{M}},\boldsymbol{M}]=0$. There are three solutions for Eqs. \eqref{omega_d_1} and \eqref{omega_d_2}:
\begin{description}
\item[$1^{st}$] $\boldsymbol{D}=0$, so $\boldsymbol{M}=0$; FLRW space-time,
\item[$2^{nd}$] $\boldsymbol{\Omega}=0$, so $\boldsymbol{O}=const.$ matrix. i.e. $\boldsymbol{M}$ can be diagonalized by a constant matrix $\boldsymbol{O}$ at any constant time hyper-surfaces.
\item[$3^{rd}$] Each of $\boldsymbol{\Omega}$ and $\boldsymbol{D}$ has only one degree of freedom:
\begin{equation}
~~~~~~~\boldsymbol{\Omega} = \left(
\begin{array}{ccc}
0 & \omega & 0 \\
-\omega & 0 & 0 \\
0 & 0 & 0
\end{array} \right) \, ,~~~~~~~~~~
\boldsymbol{D} = \left(
\begin{array}{ccc}
d & 0 & 0 \\
0 & d & 0 \\
0 & 0 & -2d
\end{array} \right) \, .
\end{equation}
So in this case $\boldsymbol{M}$ has two degrees of freedom and the space-time would be a Bianchi-{\it I} with planar symmetry. This is the case has been studied in \cite{Abolhasani:2013zya}.
\end{description}


\section{Nonlinear Gauge Transformations}
\label{nonlinear}
In this appendix, we investigate the behavior of $\mathcal{N}$ and $\boldsymbol{\mathcal{M}}$ under coordinate transformations. We define nonlinear gauge transformation by $T$ as temporal generator and $L^{i}$ as spatial shift generator,
\begin{align}
t \rightarrow \tilde{t} &= t + T(t,x^{i}) \, ,
\\
x^{i} \rightarrow \tilde{x}^{i} &= x^{i} + L^{i}(t,x^{i}) \, ,
\label{Transformation}
\end{align} 
or conversely
\begin{align}
t &= \tilde{t} + \tilde{T}(\tilde{t},\tilde{x}^{i}) \, , \\
x^{i} &= \tilde{x}^{i} + \tilde{L}^{i}(\tilde{t},\tilde{x}^{i}) \, .
\end{align} 
Under any change of the coordinates, the line element remains invariant,
\begin{align}
\mathrm{d}s^2 &= -(\alpha^2 - \beta_{k}\beta^{k})\,\mathrm{d}t^2
+2\,\beta_{i}\,\mathrm{d}t\,\mathrm{d}x^{i}
+\gamma_{ij}\,\mathrm{d}x^{i}\,\mathrm{d}x^{j}\nonumber\\
&= -(\tilde{\alpha}^2 - \tilde{\beta}_{k}\tilde{\beta}^{k})\,\mathrm{d}\tilde{t}^2
+2\,\tilde{\beta}_{i}\,\mathrm{d}\tilde{t}\,\mathrm{d}\tilde{x}^{i}
+\tilde{\gamma}_{ij}\,\mathrm{d}\tilde{x}^{i}\,\mathrm{d}\tilde{x}^{j} \, .
\end{align}
Equating the coefficient of $\mathrm{d}\tilde{t}^2$, $\mathrm{d}\tilde{t}\,\mathrm{d}\tilde{x}^{i}$, and $\mathrm{d}\tilde{x}^{i}\,\mathrm{d}\tilde{x}^{j}$ on both sides of the above, we obtain
\begin{align}
&\tilde{\alpha}^2 - \tilde{\beta}_{k}\tilde{\beta}^{k} = (\alpha^2 - \beta_{k}\beta^{k})(1+\partial_{\tilde{t}}\tilde{T})^{2}-2\,\beta_{i}(\partial_{\tilde{t}}\tilde{L}^{i})(1+\partial_{\tilde{t}}\tilde{T})-\gamma_{ij}(\partial_{\tilde{t}}\tilde{L}^{i})(\partial_{\tilde{t}}\tilde{L}^{j}) \, ,
\label{alpha}\\
&\tilde{\beta}_{i} = -(\alpha^2 - \beta_{k}\beta^{k}) \partial_{\tilde{i}}\tilde{T} (1+\partial_{\tilde{t}}\tilde{T})+\beta_{j}\big(\partial_{\tilde{i}}\tilde{T}\,\partial_{\tilde{t}}\tilde{L}^{j}+(1+\partial_{\tilde{t}}\tilde{T})(\delta^{j}_{i}+\partial_{\tilde{i}}\tilde{L}^{j}) \big)
\nonumber\\&~~~~~~
+\gamma_{ij}(\partial_{\tilde{t}}\tilde{L}^{j})(\delta^{j}_{i}+\partial_{\tilde{i}}\tilde{L}^{j}) \, ,
\label{beta}\\
&\tilde{\gamma}_{ij} = -(\alpha^2 - \beta_{k}\beta^{k}) (\partial_{\tilde{i}}\tilde{T})(\partial_{\tilde{j}}\tilde{T})+2\,\beta_{k}(\partial_{\tilde{i}}\tilde{T})(\delta^{k}_{j}+\partial_{\tilde{j}}\tilde{L}^{k})+\gamma_{kl}(\delta^{k}_{i}+\partial_{\tilde{i}}\tilde{L}^{k})(\delta^{l}_{j}+\partial_{\tilde{j}}\tilde{L}^{l}) \, .
\label{gamma}
\end{align}
From (\ref{beta}) and $\beta_{i} \sim \mathcal{O}(\epsilon)$, spatial shift generator $L^{i}$ is also $\mathcal{O}(\epsilon)$. Hence, \eqref{gamma} can be rewritten as
\begin{eqnarray}
\tilde{\gamma}_{ij}(\tilde{t},\tilde{x}^{i}) = \gamma_{ij}(t,x^{i})+\mathcal{O}(\epsilon^2) \, .
\label{gammauptoE2}
\end{eqnarray}
By taking the determinant of both sides, we will get
\begin{eqnarray}
\tilde{\mathcal{N}}(\tilde{t},\tilde{x}^{i}) = \mathcal{N}(t,x^{i})+\mathcal{O}(\epsilon^2) \, ,
\label{N}
\end{eqnarray}
which results in
\begin{eqnarray}
\tilde{N}(\tilde{t})+\tilde{\psi}(\tilde{t},\tilde{x}^{i}) = N(t)+\psi(t,x^{i})+\mathcal{O}(\epsilon^2) \, .
\label{Npsi}
\end{eqnarray}
From equations (\ref{gammauptoE2}) and (\ref{N}) we have
\begin{eqnarray}
\tilde{\mathcal{M}}_{ij}(\tilde{t},\tilde{x}^{i}) = \mathcal{M}_{ij}(t,x^{i})+\mathcal{O}(\epsilon^2) \, ,
\label{M}
\end{eqnarray}
and
\begin{eqnarray}
\tilde{M}_{ij}(\tilde{t})+\tilde{h}_{ij}(\tilde{t},\tilde{x}^{i}) = M_{ij}(t)+h_{ij}(t,x^{i})+\mathcal{O}(\epsilon^2) \, .
\label{Mh}
\end{eqnarray}

It is easy to see that $\mathcal{M}_{ij}$ transforms as a rank 2 tensor under purely spatial coordinate transformations
\begin{eqnarray}
\partial_{i} \tilde{x}^{k} \partial_{j} \tilde{x}^{l} \tilde{\mathcal{M}}_{kl}&=&(\delta^{k}_{i}+\partial_{\tilde{i}}\tilde{L}^{k})(\delta^{l}_{j}+\partial_{\tilde{j}}\tilde{L}^{l})\tilde{\mathcal{M}}_{kl}
\nonumber\\
&=&\tilde{\mathcal{M}}_{ij}(\tilde{t},\tilde{x}^{i})+\mathcal{O}(\epsilon^2)
\nonumber\\
&=&\mathcal{M}_{ij}(t,x^{i})+\mathcal{O}(\epsilon^2).
\end{eqnarray}

On the other hand we know $\boldsymbol{\hat{\gamma}}$ is a tensor density of weight -2/3,
\begin{align}
\hat{\gamma}_{ij}(t,\boldsymbol{x})&=J^{-2/3}~\partial_{i}\tilde{\boldsymbol{x}}^{k}~\partial_{j}\tilde{\boldsymbol{x}}^{l}~\tilde{\hat{\gamma}}_{kl}(\tilde{t},\tilde{\boldsymbol{x}}) \, ,
\end{align}
where 
\begin{align}
J=\det{[\dfrac{\partial \tilde{\boldsymbol{x}}^{i}}{\partial \boldsymbol{x}^{j}}]}=\det{[\delta^{i}_{j}+\partial_{j}\tilde{L}^{i}]}=1+\mathcal{O}(\epsilon^2) \, .
\end{align}
Therefore we recover (\ref{M}) again.

\section{Order Counting of Spatial Ricci Tensor}
\label{GEChris}
We use ADM metric (\ref{ADMmetric}) and calculate the Christoffel symbols
\begin{align}
\Gamma^{\alpha}_{\mu \nu}=\frac{1}{2}g^{\alpha\beta}(\partial_{\mu}g_{\beta\nu}+\partial_{\nu}g_{\mu\beta}-\partial_{\beta}g_{\mu\nu}) \, ,
\end{align}
which by assuming $\beta_{i} \sim \partial_{i} \sim \mathcal{O}(\epsilon)$ yields
\begin{align}
\Gamma^{0}_{00} &=\alpha\acute{}+\mathcal{O}(\epsilon^2) \, ,\\
\Gamma^{0}_{i0} &=\frac{1}{\alpha}\partial_{i}\alpha+\frac{\beta^{j}}{\alpha}\gamma\acute{}_{ij}+\mathcal{O}(\epsilon^2)\nonumber\\
&=\frac{\mathcal{N}\acute{}}{\alpha}\beta_{i}+\frac{1}{2\alpha}e^{2\mathcal{N}}\beta^{j}\hat{\gamma}\acute{}_{ij}+\mathcal{O}(\epsilon^2) \, ,\\
\Gamma^{0}_{ij} &=\frac{1}{2\alpha}\gamma\acute{}_{ij}+\mathcal{O}(\epsilon^2)\nonumber\\
&=\frac{\mathcal{N}\acute{}}{\alpha}\gamma_{ij}+\frac{1}{2\alpha}e^{2\mathcal{N}}\hat{\gamma}\acute{}_{ij}+\mathcal{O}(\epsilon^2) \, ,\\
\Gamma^{i}_{00} &=-\beta^{i}\alpha\acute{}+\frac{1}{2}\gamma^{ij}(\alpha\partial_{j}\alpha+\partial_{t}\beta_{j})+\mathcal{O}(\epsilon^2) \, ,\\
\Gamma^{i}_{j0} &=\frac{\alpha}{2}\gamma^{ik}\dot{\gamma}_{kj}+\mathcal{O}(\epsilon^2)\nonumber\\
&=\alpha\mathcal{N}\acute{}\delta^{i}_{j}+\frac{\alpha}{2}\hat{\gamma}^{ik}\hat{\gamma}\acute{}_{kj}+\mathcal{O}(\epsilon^2) \, ,\\
\Gamma^{i}_{jk} &={}^{3}\Gamma^{i}_{jk}-\frac{1}{2\alpha}\beta^{i}\gamma\acute{}_{jk}+\mathcal{O}(\epsilon^2) \, .
\end{align}
As noted before, the primes indicate derivatives with respect to proper time, $\partial_{\tau}$, and ${}^{3}\Gamma^{i}_{jk}$ and ${}^{3}\hat{\Gamma}^{i}_{jk}$ are spatial Christoffel symbols related to $\gamma_{ij}$ and $\hat{\gamma}_{ij}$, respectively,
\begin{align}
{}^{3}\Gamma^{i}_{jk}&=\frac{1}{2}\gamma^{il}(\partial_{j}\gamma_{lk}+\partial_{k}\gamma_{jl}-\partial_{l}\gamma_{jk})\nonumber\\
&={}^{3}\hat{\Gamma}^{i}_{jk}+\partial_{j}\mathcal{N}\delta^{i}_{k}+\partial_{k}\mathcal{N}\delta^{i}_{j}-
\partial_{l}\mathcal{N}\hat{\gamma}^{il}\hat{\gamma}_{jk} \, ,\\
{}^{3}\hat{\Gamma}^{i}_{jk}&=\frac{1}{2}\hat{\gamma}^{il}(\partial_{j}\hat{\gamma}_{lk}+\partial_{k}\hat{\gamma}_{jl}-\partial_{l}\hat{\gamma}_{jk}) \, .
\end{align}
It is easy to see that $\Gamma^{0}_{i0} \sim \Gamma^{i}_{00} \sim \Gamma^{i}_{jk} \sim {}^{3}\Gamma^{i}_{jk} \sim \mathcal{O}(\epsilon)$, and therefore ${}^{3}\mathcal{R} \sim {}^{3}\mathcal{R}_{ij} \sim \mathcal{O}(\epsilon^2)$.

\section{Two-parameter perturbation theory and separate universe picture}
\label{Two-parameter}

In the paper we employed two-parameter spacetime perturbations. Here the two-parameter perturbation theory is summarized and a smoothing map is introduced to describe the geometry of separate universe picture. 


In cosmological application of this theory, one starts with a FLRW spacetime, $\mathbb{M}_0$, called the background and a family of spacetime manifolds, $\mathbb{M}_{m,\delta}$ diffeomorphic to the background for describing the deviation from that. Here the indices $m$ and $\delta$ correspond to the order of smallness in shear and metric/matter fields perturbations, respectively. There is a 6-dimensional manifold foliated by this family $\mathbb{N} = \mathbb{M} \times \mathbb{R}^2$. A map between manifolds, $\mathbb{F}$, enables us to compare perturbed quantities with the corresponding unperturbed ones, just as in single parameter perturbation theory. The correspondence between the points of $\mathbb{M}_{m,\delta}$ and $\mathbb{M}_0$, usually called the gauge choice, is also assigned by this map, which is a two-parameter Abelian group of diffeomorphism
\begin{align}
\mathbb{F}_{m,\delta}:\mathbb{N} \longrightarrow \mathbb{N}.
\end{align}

The map $\mathbb{F}_{m,\delta}|_{\mathbb{M}_0}: \mathbb{M}_{0}  \longrightarrow \mathbb{M}_{m,\delta}$ satisfies the following properties:
\begin{enumerate}
\item[i)] $\mathbb{F}_{m_{1},\delta_{1}} \circ \mathbb{F}_{m_{2},\delta_{2}} =\mathbb{F}_{m_{1}+m_{2},\delta_{1}+\delta_{2}} \, , ~~~~~~~~\forall \delta_i , m_i \in \mathbb{R}.$
\item[ii)] $\mathbb{F}_{m,\delta} = \mathbb{F}_{0,\delta} \circ \mathbb{F}_{m,0} = \mathbb{F}_{m,0} \circ \mathbb{F}_{0,\delta} \, .$
\end{enumerate}

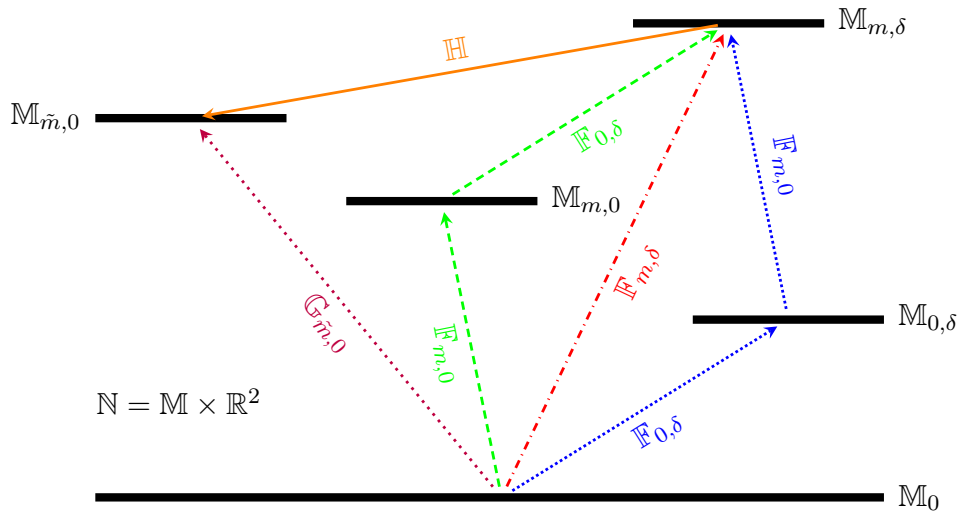
\begin{figure}
\centerline{
  \resizebox{13cm}{!}{
      \begin{tikzpicture} [scale=0.3]
     \draw [line width=3pt] (-16.5,-10) -- (16.5,-10) node [right] {$\mathbb{M}_{0}$};
     \draw [line width=3pt] (6,10) -- (14,10) node [right] {$\mathbb{M}_{m,\delta}$};;
     \draw [line width=3pt] (-6,2.5) -- (2,2.5) node [right] {$\mathbb{M}_{m,0}$};;
     \draw [line width=3pt] (8.5,-2.5) -- (16.5,-2.5) node [right] {$\mathbb{M}_{0,\delta}$};
     \draw [line width=3pt] (-8.5,6) -- (-16.5,6) node [left] {$\mathbb{M}_{\tilde{m},0}$};;
     \node (v1) at (0.5,-10) {};
     \node (v2) at (10,10) {};
     \node (v3) at (12.5,-2.5) {};
     \node (v4) at (-2,2.5) {};
     \node (v5) at (-12.5,6) {};
     \node (v6) at (-13,-6) {$\mathbb{N}=\mathbb{M} \times \mathbb{R}^2$};
     \draw[->,dashdotted, line width=1pt,shorten <=0.1pt,>=stealth,red] (v1) edge  node[sloped, below] {$\mathbb{F}_{m,\delta}$}(v2);
     \draw[->,densely dotted, line width=1pt,shorten >=0.1pt,>=stealth,blue] (v3) edge  node[sloped, above] {$\mathbb{F}_{m,0}$}(v2);
     \draw[->,densely dotted, line width=1pt,shorten >=0.1pt,>=stealth,blue] (v1) edge  node[sloped, below] {$\mathbb{F}_{0,\delta}$}(v3);
     \draw[->,densely dashed, line width=1pt,shorten >=0.1pt,>=stealth,green] (v1) edge  node[sloped, below] {$\mathbb{F}_{m,0}$}(v4);
     \draw[->,densely dashed, line width=1pt,shorten >=0.1pt,>=stealth,green] (v4) edge  node[sloped, below] {$\mathbb{F}_{0,\delta}$}(v2);
     \draw[->,dotted, line width=1pt,shorten >=0.1pt,>=stealth,purple] (v1) edge  node[sloped, below] {$\mathbb{G}_{\tilde{m},0}$}(v5);
     \draw[->, line width=1pt,shorten >=0.1pt,>=stealth,orange] (v2) edge  node[sloped, above] {$\mathbb{H}$}(v5);
     \end{tikzpicture}
  }
}
\caption{The illustration of identification maps between a FLRW background and manifolds of perturbed spacetimes. The routes shown in green (densely-dashed), red (dash-dotted) and blue (densely-dotted) are equivalent diffeomorphisms in two parameter perturbation theory. A good representative for this set of equivalence is $\mathbb{F}_{m,\delta}$. The equivalence of $\mathbb{H} \circ \mathbb{F}_{m,\delta}$ and $\mathbb{G}_{\tilde{m},0}$ up to $\mathcal{O}(\epsilon^2)$ is discussed in the paper. This picture holds for a physical spacetime (any patch in separate universe picture), $\mathbb{M}_{m,\delta}$, with its own initial conditions.}
\label{Fig1}
\end{figure}
 
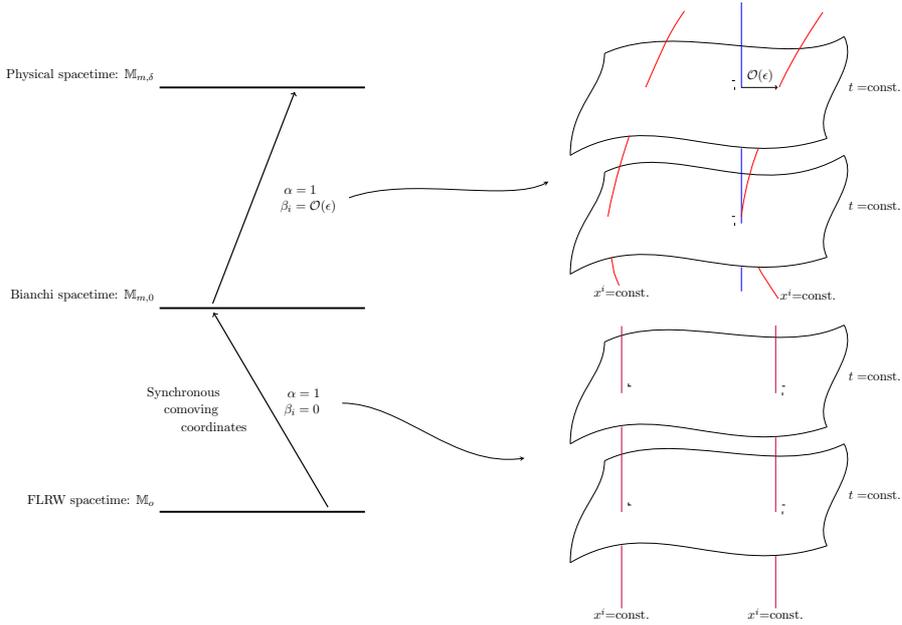
\begin{figure}
\centerline{
  \resizebox{12cm}{!}{
    \begin{tikzpicture}
            \draw [thick] (-3,2) node (v3) {} .. controls (-0.9,3) and (2.5,1.5) .. (4,2.5) node (v1) {};
            \draw [thick] (-4,-1) node (v2) {} .. controls (-1.5,0.5) and (0.5,-1.5) .. (3.5,-0.5);
            \draw [thick] (4,-1) node (v7) {} .. controls (1.5,-2.5) and (-1,-0.5) .. (-3,-1.5) node (v5) {};
            \draw [thick] (3.5,-4) node (v6) {} .. controls (0.5,-5) and (-1.5,-3) .. (-4,-4.5) node (v4) {};
            \draw [thick] (3.5,-4) .. controls (3,-3) and (4.5,-2) .. (4,-1);
            \draw [thick] (4,-6) node (v11) {} .. controls (2,-7) and (-1,-5.5) .. (-3,-6.5) node (v9) {};
            \draw [thick] (3.5,-9) node (v10) {} .. controls (0.5,-10) and (-1.5,-8) .. (-4,-9.5) node (v8) {};
            \draw [thick] (-4,-4.5) .. controls (-4,-3) and (-3,-2.5) .. (-3,-1.5);   
            \draw [thick] (4,-9.5) node (v15) {} .. controls (2,-10.5) and (-1,-9) .. (-3,-10) node (v13) {};
            \draw [thick] (3.5,-12.5) node (v14) {} .. controls (0.5,-13.5) and (-1.5,-11.5) .. (-4,-13) node (v12) {};
            \draw [thick] (3.5,-0.5) .. controls (3,0.5) and (4.5,1.5) .. (4,2.5);
            \draw [thick] (-4,-1) .. controls (-4,0.5) and (-3,1) .. (-3,2);
            \draw [ultra thick] (-10,-5.5) .. controls (-15,-5.5) and (-15.5,-5.5) .. (-16,-5.5) node [above left]{Bianchi spacetime: $\mathbb{M}_{m,0}$};
            \draw [ultra thick] (-10,1) .. controls (-15,1) and (-15.5,1) .. (-16,1) node [above left]{Physical spacetime: $\mathbb{M}_{m,\delta}$};
            \draw [ultra thick] (-10,-11.5) .. controls (-15,-11.5) and (-15.5,-11.5) .. (-16,-11.5) node [above left]{FLRW spacetime: $\mathbb{M}_o$};
            \node (v16) at (-11,-11.5) {};
            \node (v17) at (-14.5,-5.5) {};
            \draw  [->, draw=black, line width=1pt] (v16) edge node[auto]{} (v17);
            \node (v18) at (-12,1) {};
            \draw  [->, draw=black, line width=1pt] (v17) edge node[auto]{} (v18);
            \node [right] at (-12.5,-2) {$\alpha=1$};
            \node [right] at (-12.6,-2.5) {$\beta_{i}={\cal O}(\epsilon)$};
            \node [right] at (-12.4,-8) {$\alpha=1$};
            \node [right] at (-12.5,-8.5) {$\beta_{i}=0$};
            \node [right] at (-16.5,-8) {Synchronous};
            \node [right] at (-16,-8.5) {comoving};
            \node [right] at (-15.5,-9) {coordinates};
            \node [right] at (4,-11) {$t=$const.};
            \node [right] at (4,-7.5) {$t=$const.};
            \node [right] at (4,-2.5) {$t=$const.};
            \node [right] at (4,1) {$t=$const.};
            \node (v19) at (-2.5,-14.5) {$x^i$=const.};
            \node (v20) at (-2.5,-12.5) {};
            \node (v21) at (-2.5,-11.5) {};
            \node (v22) at (-2.5,-9) {};
            \draw [shorten <=-4pt,shorten >=-4pt,purple](v19) edge (v20);
            \draw [shorten <=-4pt,shorten >=-4pt,purple] (v21) edge (v22);
            \node (v23) at (-2.5,-8) {};
            \node (v24) at (-2.5,-6) {};
            \draw  [shorten <=-4pt,shorten >=-3pt,purple](v23) edge (v24);
            \node (v25) at (2,-14.5) {$x^i$=const.};
            \node (v26) at (2,-12.8) {};
            \node (v27) at (2,-11.5) {};
            \node (v28) at (2,-9.3) {};
            \node (v29) at (2,-8) {};
            \node (v30) at (2,-6) {};
            \draw [shorten <=-4pt,shorten >=-4pt,purple] (v25) edge (v26);
            \draw [shorten <=-4pt,shorten >=-3.5pt,purple] (v27) edge (v28);
            \draw [shorten <=-4pt,shorten >=-3.5pt,purple] (v29) edge (v30);
            \node (v31) at (-2.5,-7.8) {};
            \node (v32) at (-2.3,-7.8) {};
            \node (v33) at (-2.3,-8) {};
            \draw  [shorten <=4pt,shorten >=3.5pt] (v31) edge (v32);
            \draw  [shorten <=4pt,shorten >=3.5pt] (v33) edge (v32);
            \node (v34) at (-2.5,-11.3) {};
            \node (v35) at (-2.3,-11.3) {};
            \node (v36) at (-2.3,-11.5) {};
            \node (v37) at (2,-11.3) {};
            \node (v38) at (2.2,-11.3) {};
            \node (v39) at (2.2,-11.5) {};
            \node (v40) at (2,-7.8) {};
            \node (v41) at (2.2,-7.8) {};
            \node (v42) at (2.2,-8) {};
            \draw [shorten <=4pt,shorten >=3.5pt] (v34) edge (v35);
            \draw [shorten <=4pt,shorten >=3.5pt] (v36) edge (v35);
            \draw [shorten <=4pt,shorten >=3.5pt] (v37) edge (v38);
            \draw [shorten <=4pt,shorten >=3.5pt] (v38) edge (v39);
            \draw [shorten <=4pt,shorten >=3.5pt] (v40) edge (v41);
            \draw [shorten <=4pt,shorten >=3.5pt] (v41) edge (v42);
            \node (v49) at (1,-5) {};
            \node (v50) at (1,-4.3) {};
            \node (v51) at (1,-3) {};
            \node (v52) at (1,-0.8) {};
            \node (v53) at (1,1) {};
            \node (v54) at (1,3.5) {};
            \draw [shorten <=-4pt,shorten >=-4pt,blue] (v49) edge (v50);
            \draw [shorten <=-4pt,shorten >=-4pt,blue] (v51) edge (v52);
            \draw [shorten <=-4pt,shorten >=-4pt,blue] (v53) edge (v54);
            \node (v56) at (0.8,1.2) {};
            \node (v55) at (1,1.2) {};
            \node (v57) at (0.8,1) {};
            \draw [shorten <=4pt,shorten >=3.5pt] (v55) edge (v56);
            \draw [shorten <=4pt,shorten >=3.5pt] (v56) edge (v57);
            \node (v58) at (1,-2.8) {};
            \node (v59) at (0.8,-2.8) {};
            \node (v60) at (0.8,-3) {};
            \draw [shorten <=4pt,shorten >=3.5pt] (v58) edge (v59);
            \draw [shorten <=4pt,shorten >=3.5pt] (v59) edge (v60);
            \node (v61) at (1.5,-4.3) {};
            \node (v62) at (1.5,-0.8) {};
            \node (v63) at (2.1,1) {};
            \node [right] (v64) at (2,-5.1) {$x^i$=const.};
            \draw [shorten <=-4pt,shorten >=-4pt,red] (2,-5.1) .. controls (1.8,-4.8) and (1.6,-4.5) .. (v61);
            \draw [shorten <=-4pt,shorten >=-4pt,red] (v58) .. controls (1.1,-1.9) and (1.4,-1) .. (v62);
            \draw [shorten <=-4pt,shorten >=-4pt,red] (v63) .. controls (2.5,1.9) and (3.1,2.8) .. (3.3,3.1);
            \draw [->,shorten <=-4pt,shorten >=-2.5pt] (v53) edge node [above] {${\cal O}(\epsilon)$} (v63);
            \node  (v65) at (-2.5,-5) {$x^i$=const.};
            \node (v66) at (-2.8,-4) {};
            \node (v67) at (-2.9,-2.8) {};
            \node (v68) at (-2.3,-0.5) {};
            \node (v69) at (-1.8,1) {};
            \node (v70) at (-0.7,3.2) {};
            \draw [shorten <=-4pt,shorten >=-4pt,red] (v65) .. controls (-2.8,-4.3) and (-2.7,-4.5) .. (v66);
            \draw [shorten <=-4pt,shorten >=-6.5pt,red] (v67) .. controls (-2.7,-1.7) and (-2.4,-0.8) .. (v68);
            \draw [shorten <=-4pt,shorten >=-6.5pt,red] (v69) .. controls (-1.4,1.9) and (-1.2,2.5) .. (v70);
            \node (v71) at (-10.6,-2.3) {};
            \node (v72) at (-4.5,-1.7) {};
            \draw [->,thick, >=stealth] (v71) .. controls (-8.7,-1.6) and (-5.8,-2.4) .. (v72);
            \node (v73) at (-10.8,-8.3) {};
            \node (v74) at (-5.2,-9.9) {};
            \draw [->,thick, >=stealth] (v73) .. controls (-8.9,-8.3) and (-7.4,-10.2) .. (v74);
            \draw [thick] (3.5,-9) .. controls (3,-8) and (4.5,-7) .. (4,-6);
            \draw [thick] (-3,-6.5) .. controls (-3,-7.5) and (-4,-8) .. (-4,-9.5);
            \draw [thick] (-4,-13) .. controls (-4,-11.5) and (-3,-11) .. (-3,-10);
            \draw [thick] (3.5,-12.5) .. controls (3,-11.5) and (4.5,-10.5) .. (4,-9.5);
        \end{tikzpicture}
  }
}
\caption{Slicing and threading imposed on $M_{m,0}$ and $M_{m,\delta}$ by $F_{m,0}$ and
$M_{0,\delta}$, respectively.}
\label{Fig2}
\end{figure}

Since the perturbations are considered as fields living on the background, the perturbation to the background variable $T$ may pulled back to the background by $\mathbb{F}_{m,\delta}^{*}T$ and then be written in terms of the background coordinates for comparison,
\begin{align}
\delta_{\mathbb{F}}T \equiv \mathbb{F}^{\ast}_{m,\delta} T \vert_{\mathbb{M}_{0}} - T_{0} \,.
\end{align}
The picture we have in mind for the geometry of this setup is illustrated in Figure \ref{Fig1}. Each route between two submanifolds assigns a diffeomorphism which is used for identifying the points. The identification tasks are made by the vector fields that generate different gauge maps $\mathbb{F}$. These
fields are defined such that the identified points on different submanifolds, connected by map $\mathbb{F}$, have
the same coordinates $\{t,x^i\}$ and differ only in the values of coordinates $m$ and $\delta$. In our
work, the gauge maps $\mathbb{F}_{m,0}$ and $\mathbb{F}_{0,\delta}$ correspond to a specific coordinate system $\{t,x^i\}$ in $\mathbb{M}_{ m,0}$ and $\mathbb{M}_{ m,\delta}$, respectively. The time $t$ gives the slicing of the perturbed spacetime into $t=const.$ time slices (3-d spacelike hypersurfaces) and the spatial coordinates $x^i$
give the threading of the perturbed spacetime into $x^i=const.$ threads. Figure \ref{Fig2} shows the 3+1
decompositions of these manifolds. Slicings and threadings are orthogonal to each other in $\mathbb{M}_{m,0}$,
where the lapse function $\beta$ is zero.

The 2-parameter perturbation framework explained above conforms to the usual framework considered in the literature \cite{ Sopuerta:2003rg,Bruni:2002sma,Pitrou:2015iya}. The contribution of the current paper to this scenario is well described by defining a homogenization (smoothing) map, $\mathbb{H}$, such that 
\begin{align}
\mathbb{H} \lvert_{\mathbb{M}_{m,\delta}}: \mathbb{\mathbb{M}}_{m,\delta} \longrightarrow \mathbb{M}_{\tilde{m},0},
\end{align}
and we have $\mathbb{H}^2=\mathbb{H}$. This is a surjection map whose inverse image $\mathbb{H}^{-1}(p)$ is the patch to which $p$ belongs. From a mathematician point of view, the classical theory of homogenization is based on “abstract operator convergence” and deals with the asymptotic limit of a sequence of operators parameterized by a small parameter, $\delta$ for example. We, however, do not mind the details of this map, in our work. This map can smooth out the quantities on scales much larger than the anisotropy horizon $T^{-1}$. 

In Sec. 4, we showed that differential equation governing the evolution of tensor fields defined on the image of $\mathbb{H}$ are equivalent to that of those defined in $ \mathbb{M}_{m,\delta}$, in the long wavelength limit.



\end{document}